\documentclass[11pt,a4paper]{article}%

\usepackage{amsmath}%
\usepackage{amsfonts}%
\usepackage{amssymb}%
\usepackage{graphicx}%
\usepackage{graphics}%
\usepackage[sort]{natbib}%
\usepackage{breqn}%
\usepackage{setspace}
\usepackage{amsthm}
\usepackage{mathtools}
\usepackage[top=1.7in, bottom=1.7in, left=0.9in, right=0.9in]{geometry}
\usepackage{bigints}
\usepackage{epstopdf}
\usepackage{subfigure}
\usepackage{color}
\usepackage{tikz}
\usetikzlibrary{matrix,arrows,shapes,positioning,mindmap,trees,automata}
\usetikzlibrary{external}
\newcommand{\keywords}[1]{\par\addvspace\baselineskip\noindent{\em Keywords}:\enspace\ignorespaces#1}

\DeclareMathOperator\supp{supp}
\DeclareMathOperator*{\argmax}{arg\,max}

\begin{document}

\title{Bayesian model selection of vine copulas: a loss-based perspective}
\date{}
\author{Rosario Barone\thanks{Department of Statistical Sciences, Università Cattolica del Sacro Cuore, Milan, Italy}, Luciana Dalla Valle\thanks{University of Turin and University of Exeter}, Fabrizio Leisen\thanks{King's College London}, Cristiano Villa\thanks{Duke Kunshan University}}

\newtheorem{example}{\textit{Example}}
\newtheorem{definition}{Definition}
\newtheorem{theorem}{Theorem}
\newtheorem{lemma}{Lemma}
\newtheorem{proposition}{Proposition}
\newtheorem{axiom}{Axiom}
\newtheorem{remark}{\textit{Remark}}

\bibpunct{(}{)}{;}{a}{,}{,}

\maketitle

\begin{abstract}
The growing popularity of vine copulas in multivariate statistical analysis is largely driven by their ability to capture complex dependence structures. However, this flexibility comes at a cost, as the number of possible vine models grows rapidly and becomes intractable even in moderately low-dimensional settings.
These limitations affect the practical applicability of current Bayesian inference and model selection approaches, effectively restricting it to problems of relatively small-dimension due to their high computational cost.

This paper addresses the still open challenge of efficient model selection and estimation in Bayesian vine methodology. We propose a novel framework for Bayesian vine copula model selection that combines loss-based model priors with the shotgun stochastic search strategy. The strength of the proposed approach is twofold: it promotes sparsity and enables fast and effective structure selection. Furthermore, our comprehensive framework jointly identifies the vine structure, selects the copula families, and estimates the model parameters. The power of the proposed approach is demonstrated via simulation studies and an application to a real dataset of EFT portfolio asset returns.

\end{abstract}

\section{Introduction}

Vine copulas (or vines) have emerged as a flexible framework for modeling complex multivariate dependence structures \citep{bedford2001probability, bedford2002vines, aas2009pair}.
The theoretical foundation of vines is based on Sklar's theorem \citep{sklar1959fonctions}, which states that any multivariate distribution can be decomposed into its marginal distributions and a copula describing the dependence structure among variables \citep{nelsen2006introduction, joe1997multivariate}. Classical multivariate copulas can be inadequate when the dependence structure exhibits varying strengths or forms between pairs of variables. Vine copulas overcome this limitation by constructing high-dimensional dependence models through a cascade of bivariate copulas (called pair copulas), allowing to capture asymmetric relationships, tail dependence, and heterogeneous interactions among variables \citep{czado2019analyzing}.
Due to their versatility, vines have found applications in a broad range of areas, such as finance \citep{barone2023bayesian, czado2022vine}, social science \citep{dallavalle2018bayesian}, medicine \citep{michimae2023bayesian}, environmental sciences \citep{kreuzer2023bayesian}, and hydrology \citep{wu2022comparison}, to name a few.

Despite their flexibility, vine copula models present important methodological and computational challenges. The number of possible vine structures and pair copula combinations grows super-exponentially with dimension, making model selection a difficult task. 
More precisely, there exist $2^{(d-3)(d-2)/2-1} d!$ possible vine structures for $d$ variables and, for each one of them, $d(d-1)/2$ bivariate (conditional) copulas to choose from a range of different available copula families
\citep{morales2011counting}.
These drawbacks motivated substantial research efforts to develop efficient estimation and modelling strategies for vine copulas.
Early work tried to circumvent the problem of structure selection by using special
fixed vine structures, such as the canonical vine (C-vine) and the drawable vine (D-vine), both belonging to the broader class of regular vines (R-vines) \citep{aas2009pair}. However, different variable orderings still generate a high number of possible vines to choose from.
Later, \cite{dissmann2013selecting} introduced a heuristic structure selection method, which is the most widely used and remains the gold standard to this day; it is based on a greedy algorithm that utilises the absolute value of Kendall’s $\tau$. 
A similar approach in the discrete case is introduced by \cite{panagiotelis2017model}, who used greedy algorithms to automatically select vine structures and component pair copula building blocks.
A different heuristic process for structure selection using sampling orders is proposed by
\cite{zhu2020common}, which, nevertheless, hardly yields any improvement on real data compared to Dissmann's method.
More recently, \cite{vatter2025throwing} tackled the problem of vine structure selection by
proposing a heuristic method 
based on random search algorithms
and model confidence sets. However,
as the authors suggest, structure learning methods, such as the one proposed, scale efficiently in high dimension only when combined with additional sparsity-inducing mechanisms, such as truncation, regularization,
or variable selection. 

In the literature on vine copulas, a sparse model is a vine structure in which pair copula components are assumed to be independent copulas when dependence between the corresponding variables
is effectively negligible. 
Sparsity reduces model complexity while retaining the most important dependence relationships.
This stream of research is followed, for example, by \cite{muller2018representing}, who, capitalizing on the connection between vine copulas and directed acyclic graphs (DAGs), apply model selection and estimation strategies of sparse DAGs to vines. 
In a similar vein, \cite{muller2019dependence, muller2019selection} exploit the link between vines and structural equation models and induce sparsity using the Lasso. 
\cite{nagler2019model}, instead, identify sparse vine copulas via a modified version of the Bayesian information criterion (BIC). However, these contributions are available only in the frequentist framework. 

An alternative approach to induce sparsity in vine copula models is truncation, whereby all pair copulas below a certain level are set to be the independence copula. The lower the truncation level, the higher the degree of sparsity.
This strategy was first proposed by \cite{kurowicka2010optimal} specifically for the case of Gaussian copulas, utilizing partial correlations.
Truncation was later discussed by
\cite{brechmann2012truncated}, whose method involved retaining the strongest dependences in lower levels and using the Vuong's test to select an appropriate truncation depth. An extension of this method, applying Vuong's model selection tests under both nested and strictly non-nested hypotheses, was recently introduced by \cite{nishi2025model}.
A different method to determine the vine truncation depth, based on the fit indices used in the literature of structural equation models, was proposed by \cite{brechmann2015truncation}.
In a recent contribution, \cite{pfeifer2025trunc} introduced an algorithm for constructing truncated vines that could be based either on their log-likelihood or the Kullback-Leibler divergence between the distribution of the truncated vine model and the distribution of the input data.

All the aforementioned contributions to the vine methodology literature are confined to the frequentist approach. 
The only attempts to provide a full Bayesian framework for vine copula parameter estimation and model selection are proposed by \cite{gruber2015sequential, gruber2018bayesian}. Both papers are based on the reversible jump Markov chain Monte Carlo (MCMC) algorithm: the first uses a sequential model selection strategy, while the second estimates all levels of a regular vine copula simultaneously.
However, both Bayesian strategies involve an excessively high computational load and are therefore only applicable in relatively small-dimensional settings.

One of the advantages of Bayesian modelling is the ability to enforce model sparsity through an appropriate choice of prior distributions. 
\cite{gruber2015sequential, gruber2018bayesian} applied non-informative shrinkage priors, whose impact on posterior sparsity was nevertheless rather limited, while the potential of stronger sparsity-inducing priors in Bayesian vine modelling remains unexplored.

In this paper, we propose a novel loss-based prior for the selection of vine copula models, based on the ideas of \cite{villawalker2015jasa, villawalker15sjs}, who constructed objective prior distributions using the Kullback–Leibler divergence between densities of different models.
Following this approach, \cite{VillaLee2020} proposed a model prior for linear regression based on loss functions with two components: the information loss and the loss due to the model complexity. 
A similar method was used by \cite{hinoveanu2020loss} with Gaussian graphical models.
More recently, a loss-based prior was adopted for the tree topology of Bayesian Additive Regression Trees (BART) \citep{serafini2024loss}.
In the context of copulas, \cite{battagliese2023copula} employed penalized complexity priors to assess the dependence structure of bivariate copula distributions. However, the authors' work is restricted to complexity priors, to the bidimensional case and to the estimation of the copula parameter. 
In contrast, we develop a sparsity-inducing loss-based prior for model selection in multivariate regular vine copulas, thereby addressing a different, high-dimensional problem.

In order to rapidly explore the high-dimensional model space of regular vines, we developed a new Bayesian model selection approach based on the Shotgun Stochastic Search (SSS) algorithm, introduced by \cite{jones2005experiments} in the Gaussian graphical model context.
This algorithm was applied, for example, to model search in regression with large numbers of candidate predictors \citep{hans2007shotgun}, in graphical and hierarchical log-linear models \citep{dobra2010mode}, and in multiple instance learning \citep{park2024variable}.  
This method assigns higher probability to models with stronger posterior support and rapidly concentrates the search on promising regions of the model space. 

Our proposed approach allows the SSS algorithm to move between full regular vines with different orderings, truncated vines with different truncation depth, and sparse (truncated) vines with independent copulas within the same vine level.
As we show in our results, the SSS method for vine copula model selection offers a substantially more efficient and fast solution compared to currently available Bayesian alternatives.
Furthermore, implementing the SSS in parallel enhances the scalability of the proposed approach, enabling its application to high-dimensional problems and previously intractable settings. 

In summary, this paper makes a key contribution to advancing vine copula methodology.
We develop a novel fully Bayesian framework for efficient vine model selection and estimation, based on sparsity-inducing loss-based priors and scalable stochastic search techniques. We demonstrate that the proposed approach yields more parsimonious models and is computationally more tractable than state-of-the-art methods, while achieving comparable predictive performance. 

The remainder of this paper is organized as follows. 
Section 2 reviews the relevant literature and introduces the necessary background on vine copulas. Sections 3 and 4 present the proposed Bayesian framework and related computation strategy. Sections 5 and 6 report the results of simulation studies and financial data applications. Finally, Section 7 concludes.

\section{Copulas and Vines}\label{sec:copulas}

According to Sklar \citep{sklar1959fonctions}, given a vector of random variables $\textbf{X}=(X_1, \ldots, X_d)$, with $d$-dimensional joint cumulative distribution function $F(x_1, \ldots,x_d)$ and marginal cumulative distributions $F_{\delta}(x_{\delta})$, with ${\delta}=1, \ldots, d$, a $d$-dimensional copula $C$ exists, such that
$$
F(x_1, \ldots,x_d) = C(F_1(x_1), \ldots, F_d(x_d)),
$$
where $F_{\delta}(x_{\delta}) = u_{\delta}$, with $u_{\delta} \in [0,1]$. This representation is unique in the continuous case.
The joint density function can be derived by differentiation as
\begin{equation}\label{eq:joint}
f(x_1, \ldots,x_d) = c(F_1(x_1), \ldots, F_d(x_d)) \times \prod_{\delta=1}^d f_{\delta}(x_{\delta}),
\end{equation}
where $c$ denotes the $d$-variate copula density and $f_{\delta}(x_{\delta})$ denotes the marginal densities  \citep{joe1997multivariate, nelsen2006introduction}.
In an estimation context, it is common practice to adopt a two-stage approach, in which the marginal distributions are estimated in the first stage, whereas the copula is estimated in the second stage, based on the pseudo-observations, obtained as $\hat{U}_{\delta} = \hat{F}_{\delta}(X_{\delta})$, $\delta = 1, \ldots, d$ \citep{genest1995semiparametric, JoeXu1996}.

\begin{figure}[htbp]
\centering
\includegraphics[width=16cm]{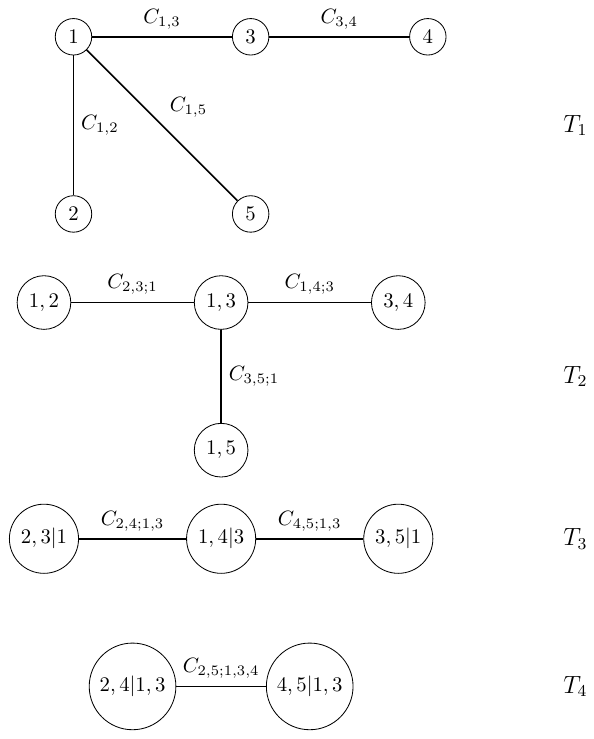}
\vspace{-8cm}\caption{R-vine tree sequence in $5$-dimensions and four trees.}\label{fig:r_vine}
\end{figure}

\begin{figure}[htbp]
\centering
\hspace{-3cm}
\includegraphics[width=19cm]{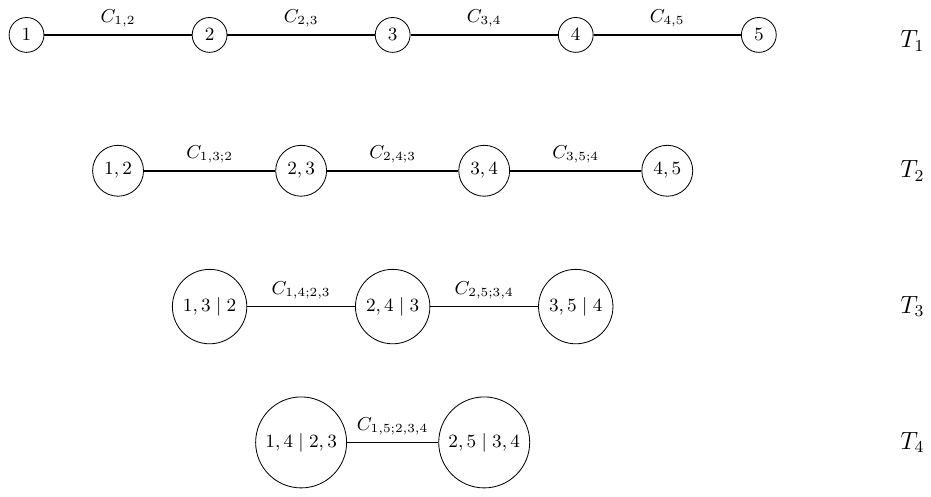}
\vspace{-13.5cm}\caption{D-vine tree sequence in $5$-dimensions and four trees.}\label{fig:d_vine}
\end{figure}

A regular vine copula (or R-vine) represents the pattern of dependence of multivariate data via a cascade of bivariate copulas, allowing us to construct flexible high-dimensional copulas using only bivariate copulas as building blocks (see \cite{joe1997multivariate, bedford2001probability, bedford2002vines, czado2019analyzing}).
The vine copula representation can be obtained factorizing the joint distribution $f(x_1, \ldots,x_d)$ of the random vector $\textbf{X}=(X_1, \ldots, X_d)$ as a product of $d(d-1)/2$ conditional densities, which can be expressed in terms of bivariate (conditional) copulas, and marginal densities.
A $d$-dimensional R-vine can be graphically represented via a nested set of trees $(T_1, \ldots, T_{d-1})$, where the variables are represented by nodes linked by edges, each associated with a pair copula. The edges of tree $T_j$ are the nodes of tree $T_{j+1}$, $j = 1, \ldots, d-1$. 
Two edges can share a node in tree $T_j$ without the associated nodes in tree $T_{j+1}$ being connected. In an R-vine, two edges in $T_j$ which become two nodes in tree $T_{j+1}$, can only share an edge if in tree $T_j$ the edges shared a common node, but they are not necessarily connected by an edge.
If all the trees in the
vine tree structure are paths, we speak of a special structure called drawable (D-)vine.
Figures~\ref{fig:r_vine} and \ref{fig:d_vine} show, respectively, examples of a general R-vine and a special D-vine structure with $5$ variables. 

\begin{figure}[htbp]
\centering
\includegraphics[width=16cm]{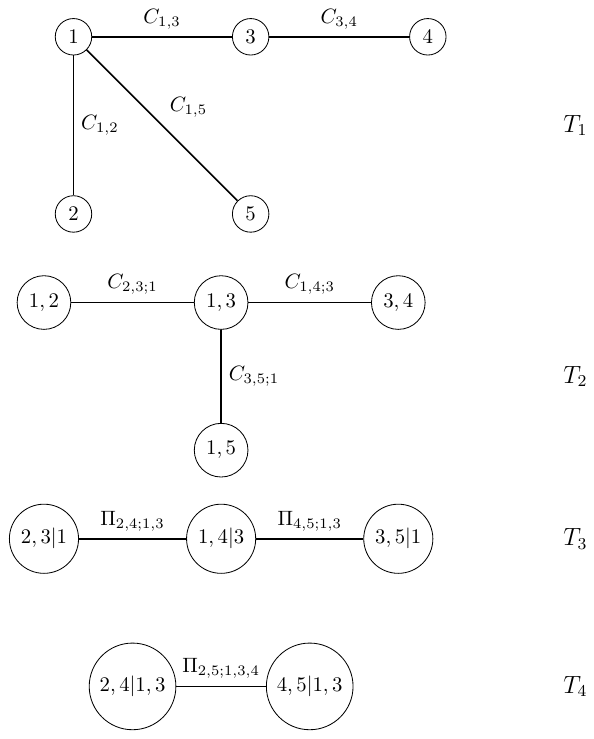}
\vspace{-8cm}\caption{R-vine copula in $5$-dimensions, truncated after the second tree $T_2$ where $\Pi_{a,b\,;\,\mathcal{D}}$ denote densities of independence copulas.
}\label{fig:r_vine_trunc}
\end{figure}

The vine copulas (such as those represented in Figures~\ref{fig:r_vine} and \ref{fig:d_vine}) identify each edge $e$ of the vine structure with a label $\{ a_e, b_e \,|\, \mathcal{D}_e \}$ and a bivariate copula $C_{a_e,\,b_e\,; \, \mathcal{D}_e}(\cdot, \cdot; \textbf{x}_{\mathcal{D}_e})$.
The sets $\{ a_e, b_e \} \subset \{1, \ldots, d\}$ and $\mathcal{D}_e \subset \{1, \ldots, d\} \backslash \{ a_e, b_e \}$ are called the conditioned set and the conditioning set, respectively \citep{CzadoNagler2022}.
Based on this, the $d$-dimensional joint density in Eq.~\eqref{eq:joint} can hence be expressed as a product of bivariate (pair) copulas and marginal distributions, using an R-vine copula representation. 
Let us consider a subrandom vector $\textbf{X}_{\mathcal{D}_e}$ of $\textbf{X}$, with value $\textbf{x}_{\mathcal{D}_e}$.
For $a_e,b_e \in \{1, \ldots, d\} \backslash \mathcal{D}_e$, we define $C_{X_{a_e},\,X_{b_e}\,; \, \textbf{X}_{\mathcal{D}_e}}(\cdot, \cdot; \textbf{x}_{\mathcal{D}_e})$ as the bivariate (pair-)copula associated with the conditional distribution of $(X_{a_e}, X_{b_e})$, given $\textbf{X}_{\mathcal{D}_e} = \textbf{x}_{\mathcal{D}_e}$, with abbreviations $C_{a_e,\,b_e\,; \, \mathcal{D}_e}(\cdot, \cdot; \textbf{x}_{\mathcal{D}_e})$ and $c_{a_e,\,b_e\,; \, \mathcal{D}_e}(\cdot, \cdot; \textbf{x}_{\mathcal{D}_e})$ for the distribution function and density, respectively.
Furthermore, $F_{X_{a_e}\,|\, \textbf{X}_{\mathcal{D}_e}}(\cdot \, | \, \textbf{x}_{\mathcal{D}_e})$ denotes the univariate conditional distribution of the random variable $X_{a_e}$, given $\textbf{X}_{\mathcal{D}_e} = \textbf{x}_{\mathcal{D}_e}$, abbreviated by $F_{a_e | \mathcal{D}_e}(\cdot | \textbf{x}_{\mathcal{D}_e})$.
Therefore, the copula density in Eq.~\eqref{eq:joint} can be written, as 
$$
c(u_1, \ldots, u_d) =  \prod_{k=1}^{d-1} \prod_{e \in E_k} c_{a_e, \, b_e \,;\,\mathcal{D}_e} \,(F_{a_e|\mathcal{D}_e}(x_{a_e}|\textbf{x}_{\mathcal{D}_e}),F_{b_e|\mathcal{D}_e}(x_{b_e}|\textbf{x}_{\mathcal{D}_e}) \, ; \, \mathbf{x}_{\mathcal{D}_e}), 
$$
where $u_{\delta} = \hat{F}_{\delta}(x_{\delta})$, $\delta = 1, \ldots, d$ \citep{czado2019analyzing}.
The simplifying assumption, that we adopt in the present work, implies that $c_{a_e,b_e\,;\,\mathcal{D}_e}$ is independent of the conditioning value $\mathbf{X}_{\mathcal{D}_e}$ \citep{Nagler2025}.
Each pair copula of the vine representation may belong to a (potentially different) parametric family, such as the Gaussian, Clayton, Gumbel and their rotated versions allowing both positive and negative dependence (see \cite{nelsen2006introduction} and \cite{joe1997multivariate}).
In a $j$-truncated vine copula, $j=1, \ldots, d-1$, all pair-copula families beyond tree level $j$ are specified as independence copulas.
Figure~\ref{fig:r_vine_trunc} depicts a
R-vine copula in $5$-dimensions, truncated after the second tree $T_2$ where $\Pi_{a,b\,;\,\mathcal{D}}$ denote densities of independence copulas.

\section{Methodology}\label{sec:methodology}

The methodology we propose is based on model priors developed for R-vine structures.  
Our approach follows the ideas introduced by the seminal paper of \cite{villawalker15sjs}, where the prior probability assigned to a model is derived from an objective assessment of the loss incurred if that model were excluded from the model space while being the true data-generating model. 
As in \cite{VillaLee2020}, we develop a loss-based model prior for R-vines, which accounts for information loss and, at the same time, explicitly incorporates model complexity. 
Indeed, high-dimensional R-vine models entail substantial costs due to their size and complexity. Therefore, it is reasonable to account for model complexity when assigning prior probabilities to competing models.

\subsection{R-vine Posterior Inference}\label{sub:inference}

As illustrated in the previous Section, given $d$ observed variables, an R-vine copula is represented by $j=1,\dots,(d-1)$ tree levels. Each pair copula in the vine may belong to a different parametric copula family selected from the set $\mathcal{C}$.
Let $\boldsymbol{u}=(u_1, \ldots, u_d)$ be the pseudo-observations, which can be obtained nonparametrically from the scaled empirical distribution functions of the marginals, or parametrically as the residual quantiles resulting from an appropriate marginal model. 

Let $\mathcal{F}$ denote the set of all possible permutations of parametric families $c\in\mathcal{C}$ within the set all admissible R-vine structures that can be constructed from $d$ variables. The set of candidate families also includes the independence copula. 
Therefore, the dependence structure is determined by selecting the most appropriate parametric family for each pair copula within each candidate R-vine model.

Let us denote the set of all candidate R-vine models by $\mathcal{S}(M)$, defined in the parameter space $\Theta_{\mathcal{S}(M)}$. 
We assume that the set $\mathcal{S}(M)$ includes every possible truncation level and, in addition, we allow for some pair copulas within the same tree level to be independence copulas.
The posterior probability of model $M_{j|\varphi}\in\mathcal{S}(M)$, with $\varphi\in\mathcal{F}$ and $j = 1, \ldots, d-1$, conditional on the data $\boldsymbol{u}$ is
\begin{equation}\label{eq:mod_posterior}
\pi(M_{j|\varphi}|\boldsymbol{u} ) \propto \pi(M_{j|\varphi})\mathcal{L}( \boldsymbol{u} | M_{j|\varphi} )
\end{equation}
where $\mathcal{L}( \boldsymbol{u} | M_{j|\varphi} )$ is the marginal (or integrated) likelihood of model $M_{j|\varphi}$ and $\pi(M_{j|\varphi})$ is the model prior probability. 
The marginal likelihood is
$$\mathcal{L}( \boldsymbol{u} | M_{j|\varphi} ) = \int_{\boldsymbol{\theta} \in \Theta_{M_{j|\varphi} }} \mathcal{L}( \boldsymbol{u}| \boldsymbol{\theta}, M_{j|\varphi}  ) \pi(\boldsymbol{\theta} | M_{j|\varphi} ) d\boldsymbol{\theta}
$$
where $\mathcal{L}( \boldsymbol{u}| \boldsymbol{\theta}, M_{j|\varphi}  )$ is the likelihood function under model $M_{j|\varphi}$ and $\pi(\boldsymbol{\theta} | M_{j|\varphi} )$ is the prior distribution of the copula parameter vector $\boldsymbol{\theta}$ for model $M_{j|\varphi}$ \citep{bernardo1994bayesian, jeffreys1998theory}. Note that $\supp{\{\boldsymbol{\theta}\}}$ depends on $c$, since different copula families may have parameters defined on different supports.

The marginal likelihood $\mathcal{L}( \boldsymbol{u} | M_{j|\varphi}  )$ is a high-dimensional integral that can be hard to calculate analytically. However, using a Laplace approximation it can be shown that, under some regularity conditions, $\mathcal{L}( \boldsymbol{u} | M_{j|\varphi}  )$ can be approximated by the Bayesian Information Criterion (BIC) as follows
$$
\mathcal{L}( \boldsymbol{u} | M_{j|\varphi}  ) \approx \exp \left\{ - \frac{1}{2} \mbox{BIC}(M_{j|\varphi} ) \right\}
$$
where
$$
\mbox{BIC}(M_{j|\varphi} ) = -2 \log \mathcal{L}( \boldsymbol{u}| \boldsymbol{\hat{\theta}}, M_{j|\varphi}  ) + \mbox{dim}(\boldsymbol{\hat{\theta}}) \log(n)
$$
with $\mbox{dim}(\boldsymbol{\hat{\theta}})$ denoting the dimension of $\boldsymbol{\hat{\theta}}$, $n$ denoting the sample size and $\boldsymbol{\hat{\theta}}$ denoting a measure of central tendency of the posterior distribution.

Therefore, we compute the posterior distribution of the parameters of the selected model $M_{j|\varphi}$ as:
\begin{equation}
\label{selectedM}
\pi(\boldsymbol{\theta} | \boldsymbol{u}, M_{j|\varphi} ) \propto \mathcal{L}(\boldsymbol{u} | \boldsymbol{\theta} , M_{j|\varphi} ) \pi( \boldsymbol{\theta} | M_{j|\varphi}  )
\end{equation}

where $\pi(\boldsymbol{\theta} \mid M_{j|\varphi})$ is specified as the product of independent uniform priors over the admissible parameter space of the selected copula families. Since the support of each copula parameter depends on the corresponding family, the bounds of the uniform prior are family-specific.

\subsection{Model Priors for R-Vine Structures}\label{sub:priors}

In order to determine the model prior probabilities $\pi(M_{j|\varphi})$ in Eq.~\eqref{eq:mod_posterior}, we consider the loss incurred if the model were removed from the model space while it is, in fact, the true model.
The loss will have two main components: a loss in information and a loss due to complexity.

\paragraph{Loss in information}
The loss of information is grounded in the well-known Bayesian principle (see, for example, \citet{berk1966limiting}) that, if the true model is removed from consideration, the posterior distribution asymptotically concentrates on the model that is closest to the true model in terms of Kullback--Leibler divergence (see \cite{kullback1951information} for the original divergence formulation and \cite{killiches2018model} for computationally feasible approximations in large dimension).
Intuitively, this nearest model is obtained by considering the minimum perturbation of the model under investigation. 
Keeping this in mind, the loss in information for the R-vine model $M_{j|\varphi}$, with $j=1, \ldots, d-1$, $j \neq j'$ and $\varphi \in \mathcal{F}$, can be formalised as:
$$
\mbox{Loss}_I(M_{j|\varphi}) = \min_{j\neq j'}D_{KL}\left(M_{j|\varphi}\|M_{j'|\varphi}\right).
$$
The loss of information is intrinsic to the structure itself. That is, for a fixed R-vine model (including the type of vine copula, the marginal distributions, the copula families, and the node–edge configuration), the loss is determined independently of any subjective input from the decision maker (or experimenter). Implicitly, the prior assumes that the selected vine type and structure are supported by a sound rationale and therefore takes them as given.

\paragraph{Loss due to complexity}
As we move from a highly sparse R-vine model (such as, for example, an R-vine truncated at the first tree level) to a full (``saturated'') R-vine model with no independence copulas, the model becomes increasingly complex at each tree level. It is therefore important to account for this complexity, balancing the gain in information provided by a richer model against the penalty associated with a more complex structure. This could be implemented by combining complexity factors such as 
the number of copulas in the model and the depth of the tree structure.

Therefore, we propose the following complexity loss function for the R-vine model $M_{j|\varphi}$, with $j=1, \ldots, d-1$ and $\varphi \in \mathcal{F}$:
$$
\mbox{Loss}_C(M_{j|\varphi}) = c_1\log(k_{j|\varphi}) + c_2\,l_{j|\varphi},
$$
where $c_1,c_2>0$ are tuning constants that allows to control the strength of the penalty, $k_{j|\varphi}$ denotes the number of copulas in the model, whilst $l_{j|\varphi}$ denotes the ``depth'' of the model, measured as the tree level (starting from $0$ for the independent model, where all copulas in the vine are independence copulas). 
Our proposal is motivated by the observation that the number of copulas grows nonlinearly with $d$. Therefore, the logarithm of $k_{j|\varphi}$, combined with an appropriate choice of $c_1$, allows us to achieve the desired level of sparsity in the prior. Note that we exclude the case $c_1 = 0$, as it is desirable to always include a penalty term on the number of copulas in order to properly account for model complexity.

Note that, since loss functions are defined up to a positive constant, we assume $c_1,c_2>0$. The constants allow for a great deal of flexibility in tuning the prior. By choosing relatively small values of the constants, the prior convergence to a uniform prior assigning equal mass to each model. Conversely, as the values of the constants increase, the prior probability decreases more rapidly. Therefore, different degrees of sparsity can be encouraged by appropriately specifying the prior.

Furthermore, the constants $c_1$ and $c_2$ can be calibrated to encode prior beliefs about key structural features of the model, such as the expected depth of the vine and the expected number of copulas.

\paragraph{Loss-based prior for R-vines} The prior mass proposed for model $M_{j|\varphi}$, with $j=1, \ldots, d-1$ and $\varphi \in \mathcal{F}$, is proportional to the loss of the model, given by
$$\mbox{Loss}(M_{j|\varphi}) = \mbox{Loss}_I(M_{j|\varphi})+\mbox{Loss}_C(M_{j|\varphi}),$$
resulting in the general formulation of the prior as
\begin{eqnarray}\label{eq:fullprior}
P(M_{j|\varphi}) &\propto& \exp\left\{-\mbox{Loss}(M_{j|\varphi})\right\} \nonumber \\
&=& \exp\left\{\min_{j\neq j'}D_{KL}\left(M_{j|\varphi}\|M_{j'|\varphi}\right) - c_1\log(k_j) - c_2\,l_j\right\}, \qquad c_1,c_2>0.
\end{eqnarray}

\begin{figure}[!ht]
\centering
\begin{subfigure}
  \centering
\includegraphics[width=.8\linewidth]{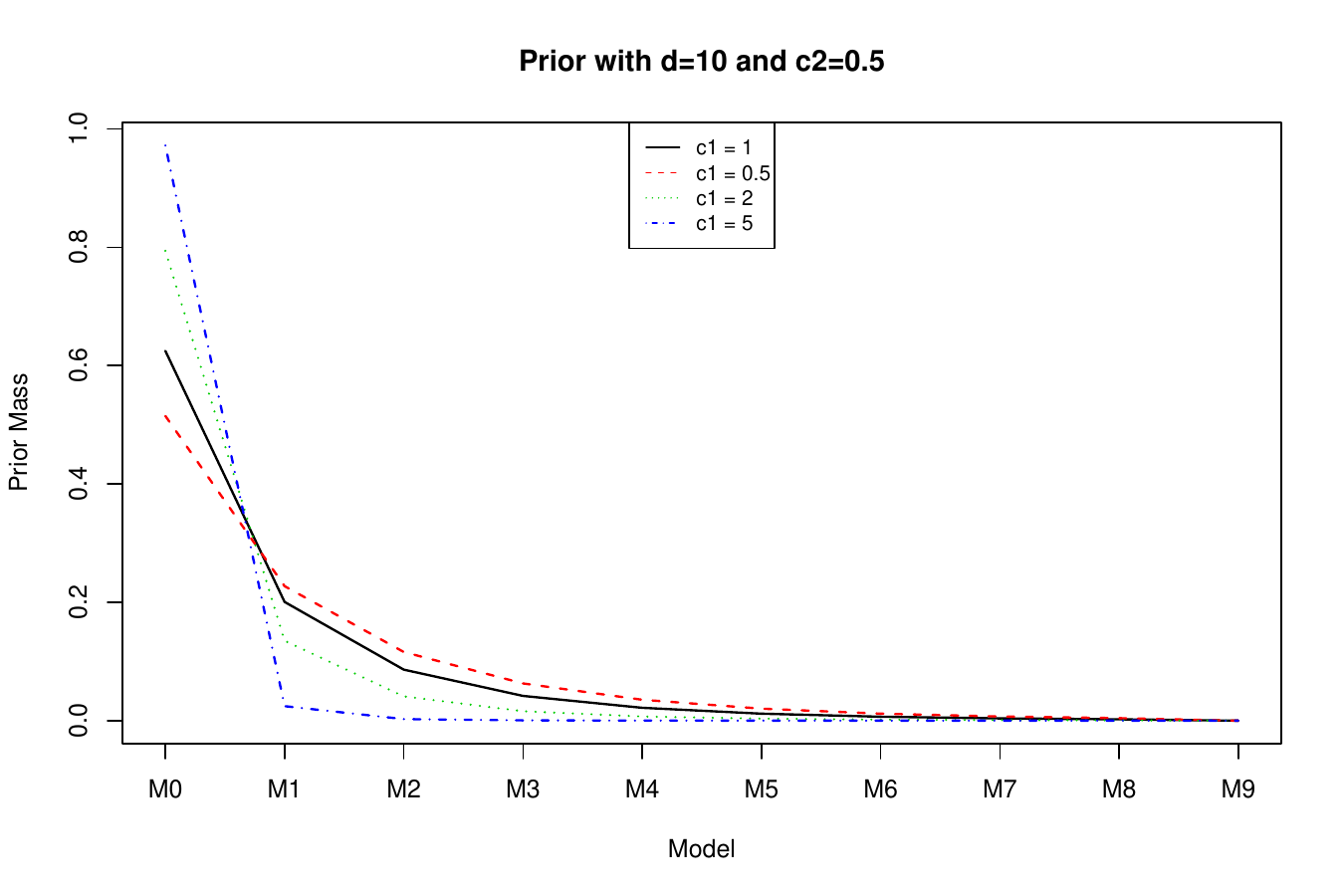}
\end{subfigure}%
\begin{subfigure}
  \centering
\includegraphics[width=.8\linewidth]{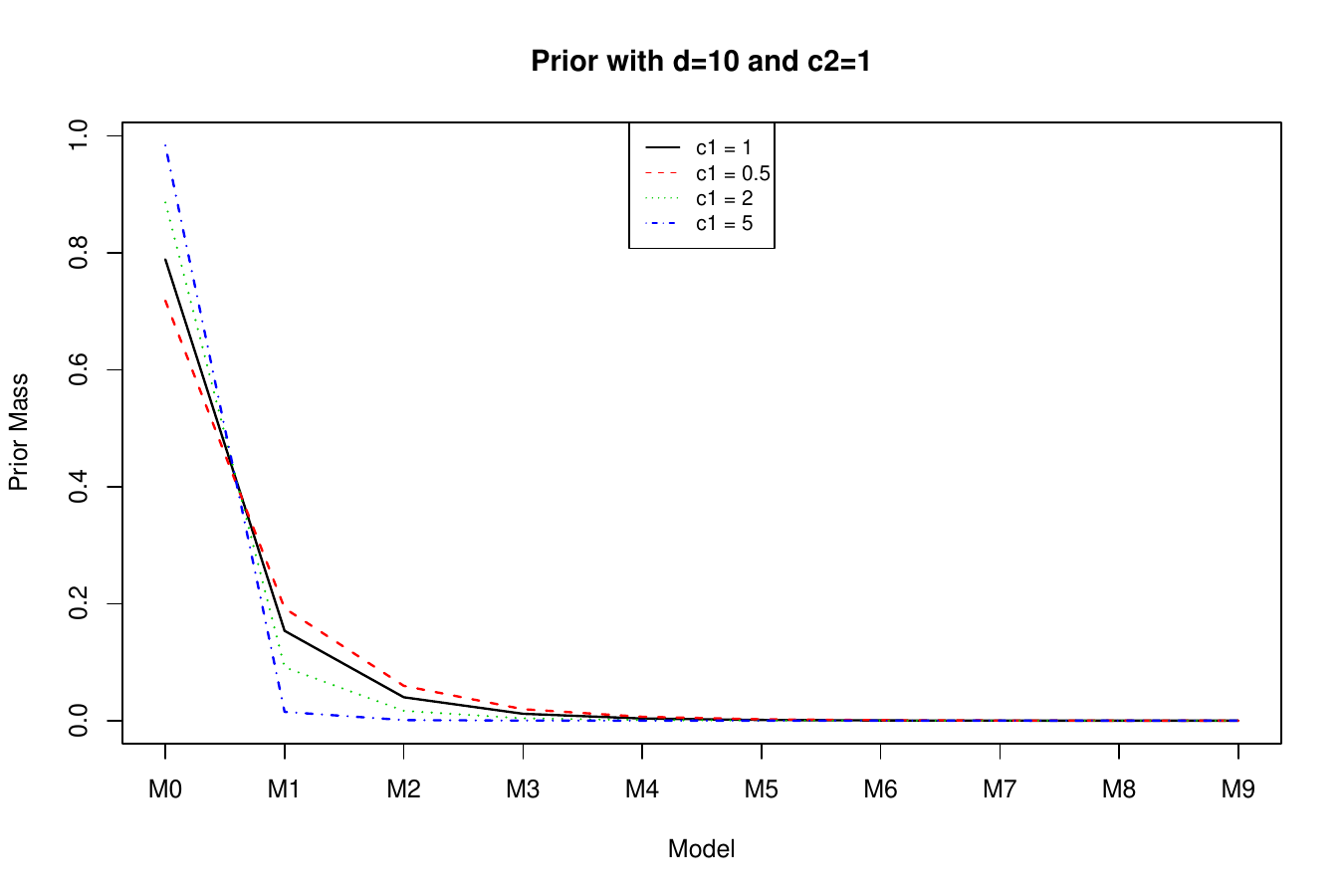}
\end{subfigure}
\caption{Behaviour of the prior, on the $10$ different truncated models that can be obtained by a D-vine with $d=10$, for different values of $c_1$ and $c_2$.
$M_0$ denotes the independence model, whilst $M_9$ denotes the full D-vine structure.}\label{fg:fullprior_d10}
\end{figure}

Figure~\ref{fg:fullprior_d10} illustrates the behaviour of the prior in Eq.~\eqref{eq:fullprior}, on the $10$ different truncated models that can be obtained by a D-vine with $d=10$, for different values of $c_1$ and $c_2$.
$M_0$ denotes the independence model, whilst $M_9$ denotes the full (``saturated'') D-vine structure. We note that the higher the values of $c_1$ and $c_2$ the stronger the penalty for complex vine structure and the higher the sparsity induced by the proposed prior on the model. 

In Appendix~\ref{se:KL}, we derive the Kullback–Leibler divergence and the related loss of information prior mass for vine-based models in which as the pair-copulas as well as the marginal distributions belong
to the Gaussian family.

\section{Posterior Sampling}
\label{sc_modelselection}

Posterior sampling is performed developing a parallelizable SSS algorithm for rapid traversal of spaces of the R-vine.

The proposed computational strategy involves a two-phases estimation of the posterior distributions:
\begin{enumerate}
    \item Definition of the optimal model, considering the truncation level of the vine and the parametric families associated with each pair copula.
    \item Estimation of the posterior distributions of the selected model.
\end{enumerate}
The first phase involves exploring the discrete space of models using a SSS algorithm, which is a method that allows efficient movement within the vine model space. 
The idea of \cite{jones2005experiments} is to generate, at each step, a large number of candidate models, “shooting out”
neighbour candidates in all directions and then selecting plausible candidates based on posterior probabilities.
Convergence is fairly rapid. 
At each step of the SSS, the estimation of a central tendency value of the parameter posteriors for all evaluated models is computed, which will be used in the model selection phase. 
The second phase involves estimating the posterior of the parameters of the chosen model, which is implemented using MCMC.


\subsection{MAP-Based Shotgun Stochastic Search for R-vine Models}

The motivation for the development of a SSS-based strategy is the following.
If dimension is relatively small, we may consider implementing an additional Metropolis-Hastings step for the selection of the vine structure.  However, as dimension increases, we may face the challenge of computing the posterior probability for each possible vine structure, leading to a considerable slowdown in computations. 
To overcome this drawback, we develop a computational strategy based on the SSS algorithm \citep{jones2005experiments}, which assumes that the search is an MCMC tool allowing us to visit elements in the model space.  The algorithm is outlined as follows:
\begin{enumerate}
\item \textbf{Initialization}:  We start with a generic structure, say $M_{j|\varphi}$, with truncation level $j=1,\dots,d-1$ and a combination of pair copula families $\varphi\in\mathcal{F}$.
\item \textbf{Model selection}: We create a subset of the model space $\mathcal{B}(M_{j|\varphi})$, that is a neighborhood of $M_{j|\varphi}$, obtained by considering models that differ from $M_{j|\varphi}$ by only one element, meaning they have one extra (or one less) copula in the R-vine tree. 
We identify the optimal model, that is, the combination of the optimal tree level together with the optimal copula families, by maximizing the discrete posterior distribution $\pi(M_{j|\varphi}| \boldsymbol{u})$ defined on the model space $\mathcal{B}(M_{j|\varphi})$, that is
\begin{equation}
\label{MAP_model}
\hat{M}_{MAP}=\argmax_{M_{j|\varphi}\in \mathcal{B}(M_{j|\varphi})}\pi(M_{j|\varphi}| \boldsymbol{u}).
\end{equation}
In other words, after building up the set $\mathcal{B}(M_{j|\varphi})$ of cardinality $N_1$, the maximization in Eq.~\eqref{MAP_model} is performed in two phases:
\begin{enumerate}
\item \textbf{Family selection}: for each of the $N_1$ structures in $\mathcal{B}(M_{j|\varphi})$, we calculate
the maximum a posteriori under $M_{j|\varphi}$, denoted by $\hat{\boldsymbol{\theta}}_{MAP}$, as:
$$
\hat{\boldsymbol{\theta}}_{MAP} = \operatorname*{arg\,max}_{\theta} \pi(\boldsymbol{\theta} | \boldsymbol{u}, M_{j|\varphi} ), 
$$
from the posterior distribution in Eq.~\eqref{selectedM}.
Then, we select the optimal families $\boldsymbol{\ell} \in \mathcal{F}$ that maximize the posterior distribution for each considered structure. This step is performed in parallel.
\item \textbf{Structure selection}: Among the resulting $N_1$ optimal structures, we select the R-vine structure $M_{k|\ell}$, with $k=1,\dots,(d-1)$ and $\ell\in\mathcal{F}$ based on the posterior probabilities. 
\end{enumerate}
The selected structure $M_{k|\ell}$, that is equivalent to the maximum a posteriori defined in \eqref{MAP_model}, is taken as a new starting point.
\item Return to step 2.
\end{enumerate}

Step 2 might seem computationally intensive. Indeed, it involves exploring a very large parametric space, considering not only the optimal structures but also the various parametric families within $\mathcal{C}$ for each pair copula. However, the SSS approach accelerates computation by exploring a subset of the model space at each iteration. The update of the parameter set of the R-vine structure for each proposed family is performed by calculating the MAP exclusively for the parameters of the proposed pair copula, considering the parameters of the initial structure $M_{j|\varphi}$ as known. This approach allows us to reduce the originally multi-dimensional optimization problem to a uni-dimensional one, making computation manageable even in contexts with a large number of variables $d$. Finally, the parallelization of this step makes the computation fast and convergence is achieved very rapidly.  

\subsection{MCMC Update for the Selected Model Parameters}

Once the MAP-SSS has reached convergence, we derive the posterior distribution of the model parameters of the selected mode $M_{k|\ell}$ as in Eq.~\eqref{selectedM}. We draw samples from this posterior distribution using a Metropolis-Hastings (M-H) algorithm with proposal  
$$
q(\boldsymbol{\theta} \rightarrow \boldsymbol{\theta^*}) \sim \mathcal{TN}(\boldsymbol{\theta}, \Sigma),
$$
where $\mathcal{TN}$ is a multivariate normal truncated in the boundaries $({a},{b})$ with covariance matrix $\Sigma$. Obviously, the vectors defining the boundaries of the truncation interval ({$a$} and {$b$}) have the same length as the number of pair copulas considered in the structure, while their specific values may vary based on the support of the elements of $\boldsymbol{\theta}$ and, consequently, on the parametric families considered within the vine. To enhance parameter convergence, we employ an adaptive scheme in which the covariance of the proposal, $\Sigma$, is updated by calculating the covariance matrix based on the accepted values of $\boldsymbol{\theta}$. The acceptance probability for this M-H step is therefore $$\alpha = \min \left\{ 1, \frac{\mathcal{L}({u} | \boldsymbol{\theta^*} , {M_{k|\ell}} ) \pi( \boldsymbol{\theta^*} | {M_{k|\ell}}  ) q(\boldsymbol{\theta^*} \rightarrow \boldsymbol{\theta})}{\mathcal{L}({u} | \boldsymbol{\theta} , {M_{k|\ell}} ) \pi( \boldsymbol{\theta} | {M_{k|\ell}}  )q(\boldsymbol{\theta} \rightarrow \boldsymbol{\theta^*})} \right\}.
$$

\section{Simulation Study}

In this Section, we present a detailed simulation study, divided into three parts, aiming to numerically validate the properties of the proposed approach. Specifically, our primary objective is to demonstrate that the proposed method has a tolerable error rate for selecting the copula families of the vine, which decreases as the sample size increases. Subsequently, we assess the sensitivity of the proposed method to varying values of the constants $c_1$ and $c_2$ of the loss-based prior defined in Eq.~\eqref{eq:fullprior}.

\subsection{Validation of Performance}\label{sub:simul_1}

In the first part of the simulation study we test i) the ability of the proposed approach to determine the optimal number of levels in a R-vine copula; ii) its capability to select the appropriate copula families for modeling the dependence between variables. More specifically, in Tables~\ref{M1}-\ref{M4} we show the results of the implementation of the proposed method in two scenarios: a $4$-dimensional case and a $6$-dimensional case with sample size $n=(50,100)$, respectively. 
For each scenario, $100$ samples were generated from D-vine models with orderings $X_1 - X_2 - \ldots - X_4$ and $X_1 - X_2 - \ldots - X_6$, respectively. 
We assume that the models available in both scenarios are given by all possible truncations, resulting in models $M_0$ to $M_{d-1}$, with $d=4$ and $d=6$ respectively, as listed below:
\begin{eqnarray*}
M_0: && \left\{f_1\cdots f_d\right\} \\
M_1: && \left\{c_{1,2}c_{2,3}\cdots c_{d-1,d} \, \times f_1 \cdots f_d  \right\} \\
\cdots && \cdots \\
M_{d-1}: && \left\{ c_{1,2}c_{2,3}\cdots c_{d-1,d} \cdots \,c_{1,d|2,\ldots,d-1}  \, f_1\cdots f_d\right\}.
\end{eqnarray*}
For each pair-copula, we report the misclassification rate in the selection of the copula family. 
Details about the structures and parametric families used to generate the data are reported in Tables~\ref{M1} and \ref{M3} for the $4$- and $6$-dimensional case, respectively. 
In both scenarios, we assume that the true model is D-vine truncated at the first tree level, that is a model of type $M_1$. 
As summary statistics, we report the posterior estimates of Kendall's $\tau$, a nonparametric measure of dependence. This quantity allows direct comparison between the true and estimated dependence structures and enables the calculation of distance measures even when the selected pair-copula family differs from the data-generating family. 
Notably, it can be observed that the expected values of the estimated Kendall's $\tau$ are close to the data generating values of $\tau$, and all credible intervals calculated for the estimated $\tau$ contain the data generating values. 
Tables \ref{M1} and \ref{M3} show that, as the sample size increases, both the error rate in pair-copula family selection and the Mean Squared Error (MSE) of the estimated Kendall's $\tau$ decrease.
Tables~\ref{M2} and \ref{M4}, on the other hand, show the frequency distributions of models $M_0$ to $M_{d-1}$, with $d=4$ and $d=6$, respectively.
From the Tables we note that the estimated truncation levels are highly concentrated on the data generating model, already with $n=50$.
\begin{table}[ht]
\centering
\scalebox{0.8}{
\begin{tabular}{|l|rrr|}
  \hline
  \textbf{Data generating family} & \textbf{Gaussian} & \textbf{Clayton} & \textbf{Rot. Gumbel $90^o$} \\
  \hline
  \multicolumn{4}{|l|}{$n=50$} \\
  \hline
  Selection error rate &  0.34  & 0.40   & 0.81 \\
    \hline
 & $\tau_{12}$ & $\tau_{23}$ & $\tau_{34}$ \\
  \hline
$E(\cdot|X)$ & 0.36 & 0.31 & -0.51 \\ 
$SD(\cdot|X)$ & 0.07 & 0.06 & 0.06 \\ 
$q_{0.025}(\cdot|X)$  & 0.23 & 0.21 & -0.60 \\ 
$q_{0.975}(\cdot|X)$ & 0.48 & 0.44 & -0.38 \\ 
$MSE(\cdot)$ & 0.03 & 0.02 & 0.03 \\ 
  \hline
  \multicolumn{4}{|l|}{$n=100$} \\
  \hline
  Selection error rate &   0.26  & 0.21   & 0.41 \\
    \hline
 & $\tau_{12}$ & $\tau_{23}$ & $\tau_{34}$ \\
  \hline
$E(\cdot|X)$ & 0.34 & 0.30 & -0.50 \\ 
$SD(\cdot|X)$ & 0.06 & 0.04 & 0.04 \\ 
$q_{0.025}(\cdot|X)$ & 0.22 & 0.22 & -0.57 \\ 
$q_{0.975}(\cdot|X)$ & 0.44 & 0.37 & -0.42 \\ 
$MSE(\cdot)$ & 0.02 & 0.01 & 0.01 \\ 
  \hline
\end{tabular}
}
\caption{Simulation study: summary statistics of the posterior estimates for 100 samples generated from a level-one truncated ($M_1$) D-vine copula structure with $d=4$ and $n=(50,100)$. 
The label “Rot. Gumbel $90^\circ$” denotes a Gumbel copula rotated by $90^\circ$, allowing for negative dependence. 
The subscripts in $\tau_{ab}$ refer to the pair of variables involved in the corresponding pair copula, while the reported values are posterior summaries of Kendall’s $\tau$, used to compare dependence across different copula families.}
\label{M1}
\vspace{0.3cm}
\begin{tabular}{|l|rrrr|}
\hline
  &$M_0$   & $\boldsymbol{M_1}$ & $M_2$ & $M_3$  \\
 \hline
 $n=50$ &  0.00   & {0.98}  & 0.02  &   0.00    \\
  \hline
  $n=100$&  0.00   & {0.98}  & 0.02  &   0.00   \\
  \hline
\end{tabular}
    \caption{Simulation study: frequencies of the selected truncation level across 100 samples with $d=4$ and $n=(50,100)$. Bold text indicates the column corresponding to the data generating model.}
\label{M2}
\end{table}

\begin{table}[ht]
\centering
\scalebox{0.8}{
\begin{tabular}{|l|rrrrr|}
  \hline
  \textbf{Data generating family} & \textbf{Gaussian} & \textbf{Gaussian}   & \textbf{Clayton}  & \textbf{Rot. Clayton $90^o$} & \textbf{Rot. Gumbel $90^o$} \\
    \hline
    \multicolumn{6}{|l|}{$n=50$} \\
  \hline
Selection error rate & 0.33 & 0.19 &  0.23 & 0.57 & 0.73 \\
  \hline
 & $\tau_{12}$ & $\tau_{23}$ & $\tau_{34}$ & $\tau_{45}$ & $\tau_{56}$ \\ 
  \hline
$E(\cdot|X)$ & 0.34 & -0.34 & 0.35 & -0.33 & -0.35 \\ 
$SD(\cdot|X)$ & 0.07 & 0.07 & 0.07 & 0.06 & 0.07 \\ 
$q_{0.025}(\cdot|X)$ & 0.20 & -0.47 & 0.22 & -0.48 & -0.48 \\ 
$q_{0.975}(\cdot|X)$ & 0.47 & -0.22 & 0.45 & -0.23 & -0.21 \\ 
$MSE(\cdot)$& 0.02 & 0.02 & 0.01 & 0.02 & 0.02 \\ 
  \hline
  \multicolumn{6}{|l|}{$n=100$} \\
    \hline
 Selection error rate & 0.22 & 0.14 &  0.11 & 0.19  & 0.53 \\
  \hline
 & $\tau_{12}$ & $\tau_{23}$ & $\tau_{34}$ & $\tau_{45}$ & $\tau_{56}$ \\ 
  \hline
$E(\cdot|X)$ & 0.34 & -0.34 & 0.34 & -0.32 & -0.34 \\ 
$SD(\cdot|X)$  & 0.05 & 0.05 & 0.05 & 0.04 & 0.05 \\ 
$q_{0.025}(\cdot|X)$  & 0.23 & -0.43 & 0.26 & -0.40 & -0.43 \\ 
$q_{0.975}(\cdot|X)$  & 0.42 & -0.24 & 0.43 & -0.23 & -0.24 \\ 
$MSE(\cdot)$ & 0.01 & 0.01 & 0.01 & 0.01 & 0.01 \\ 
   \hline
\end{tabular}
}
\caption{Simulation study: summary statistics of the posterior estimates for 100 samples generated from a level-one truncated ($M_1$) D-vine copula structure with $d=6$ and $n=(50,100)$. 
The label “Rot. Clayton (Gumbel) $90^\circ$” denotes a Clayton (Gumbel) copula rotated by $90^\circ$, allowing for negative dependence. 
The subscripts in $\tau_{ab}$ refer to the pair of variables involved in the corresponding pair copula, while the reported values are posterior summaries of Kendall’s $\tau$, used to compare dependence across different copula families.}
\label{M3}
\vspace{0.3cm}
\begin{tabular}{|l|rrrrrr|}
\hline
  &$M_0$   & $\boldsymbol{M_1}$ & $M_2$ & $M_3$ & $M_4$ & $M_5$  \\
 \hline
 $n=50$ &  0.00   & {0.79}  & 0.21  &   0.00   &   0.00   &   0.00  \\
  \hline
  $n=100$&  0.00   & {0.82}  & 0.16  &   0.02    &   0.00   &   0.00  \\
  \hline
\end{tabular}
    \caption{Simulation study: frequencies of the selected truncation level across 100 samples with $d=6$ and $n=(50,100)$. Bold text indicates the column corresponding to the data generating model.}
\label{M4}
\end{table}

\subsection{Consistency of Model Selection}\label{sub:simul_2}

\begin{figure}
    \centering
    \includegraphics[scale=0.6]{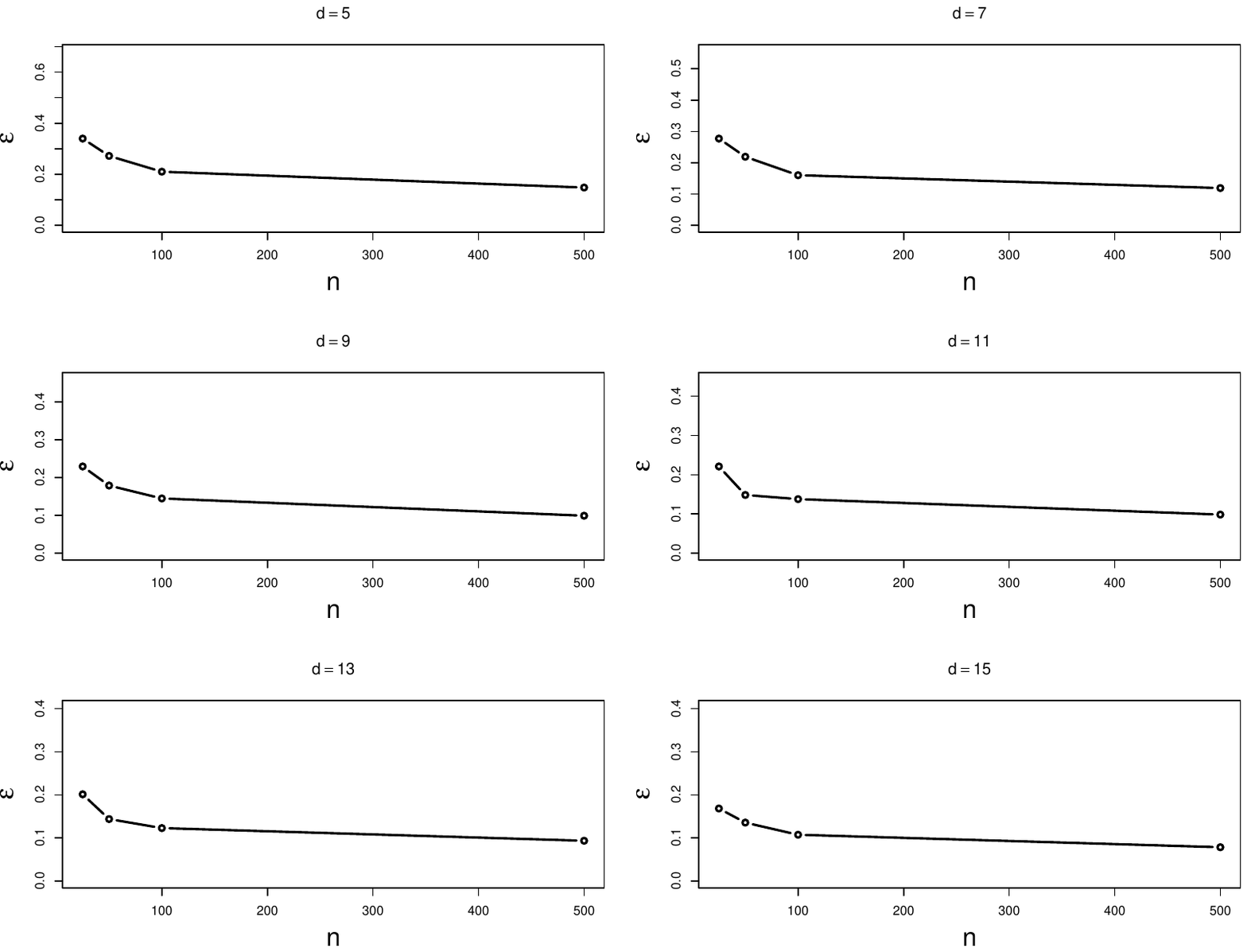}
    \caption{Simulation study: error rate $\epsilon$ ($y$-axis) against sample size $n$ ($x$-axis) as the dimension $d$ varies (in different panels), for model selection assessment.}
    \label{mse.trend}
\end{figure}

In the second part of the simulations, we implement a broader study to assess the consistency of the model selection of the proposed approach. 
Specifically, our simulation framework involves generating random R-vines with varying dimensions ranging from $d=5$ to $d=15$. For each dimension, we generate 100 random R-vine structures, with random truncation level between $d-1$ and $d-3$ and sample sizes $n = 25, 50, 75, 100, 500$. Let $\epsilon$ denote the overall model selection error rate, defined as the proportion of incorrectly selected pair-copula models relative to the total number of pair-copulas in the vine.
We implemented the proposed approach, estimating the unknown copula families along with the copula parameters, demonstrating that $\epsilon$ decreases as the sample size $n$ increases, as shown in Figure \ref{mse.trend} and Table \ref{TabError}. We observe a clear and systematic decrease in the error rate $\epsilon$ as the sample size $n$ increases, across all considered dimensions. 
This result confirms the consistency of the proposed model selection procedure: as more information becomes available, the method becomes increasingly effective at correctly identifying both the copula families and the truncation level.
Moreover, the results show that the error rate does not deteriorate with the dimensionality of the problem. On the contrary, for larger values of $d$, the error is comparable or even smaller, suggesting that the method scales well in moderately high dimensions. 
This behaviour can be explained by the fact that, although the number of pair copulas increases with $d$, the signal contained in the data also grows, allowing for more accurate estimation of the dependence structure.
Finally, the reduction in $\epsilon$ is particularly pronounced when moving from small to moderate sample sizes (e.g. from $n=25$ to $n=100$), while improvements become more gradual for larger samples. This indicates that the proposed approach is already reliable for moderate sample sizes and becomes increasingly accurate as $n$ grows.

\begin{table}[ht]
\centering
\begin{tabular}{|l|r|r|r|r|}
  \hline
 & $n=25$ & $n=50$ & $n=100$ & $n=500$ \\ 
  \hline
$d=5$ & 0.32 & 0.26 & 0.20 & 0.14 \\ 
$d=7$ & 0.27 & 0.24 & 0.18 & 0.14 \\ 
$d=9$ & 0.22 & 0.16 & 0.15 & 0.12 \\ 
$d=11$ & 0.22 & 0.15 & 0.13 & 0.09 \\ 
$d=13$ & 0.20 & 0.14 & 0.10 & 0.07 \\ 
$d=15$ & 0.17 & 0.14 & 0.11 & 0.07 \\ 
   \hline
\end{tabular}
\caption{Simulation study: error rates as the sample size $n$ (columns) and the dimension $d$ (rows) vary for model selection assessment.}
\label{TabError}
\end{table}

\subsection{Prior Sensitivity Analysis}\label{sub:simul_3}

In the final part of the simulation study, we assess the performance of the proposed method in estimating the truncation level of R-vine models as the parameter $c_1$, of the prior in Eq.~\eqref{eq:fullprior} varies, with $c_2=(0.5,1)$. 
We generate observations from random R-vine copula structures with $d=6$ and $d=8$, considering various data-generating truncation levels. Specifically, using a similar notation as in Section~\ref{sub:simul_1}, data is generated from model $M_1$ (truncated at the first tree) to $M_5$, for $d=6$, or to $M_7$ (full or ``saturated'' vine), for $d=8$. We consider the values $c_1=(0.1, 1, 5)$ and $c_2=(0.5, 1)$ and report the frequencies of the estimated truncation levels across 100 samples for each considered level.
Tables~\ref{Sensitivity1} and \ref{Sensitivity2} show the frequencies of the selected truncation levels across 100 samples generated from a random R-vine copula structure for each combination of $c_1$ and $c_2$, with $d=6$, for $n=25$ and $n=100$, respectively.
Tables~\ref{Sensitivity3} and \ref{Sensitivity4} present similar results for $d=8$, again for $n=25$ and $n=100$, respectively.
In general, with smaller sample sizes, there is clear evidence of the priors influencing the selected truncation level, with stronger effects as the size of the vine copula grows. For both $d=6$ and $d=8$, as $c_1$ and/or $c_2$ increase, there is a tendency to select lower truncation levels, including $M_0$, where all pair copulas of the R-vine are independence copulas.
In particular, the results highlight that for $n=25$, the effect of the prior is quite pronounced, with higher values of $c_1$ and $c_2$ leading to more parsimonious estimates and sparse model. This effect becomes more evident at higher truncation levels, as seen in Table \ref{Sensitivity3}, where true models $M_6$ or $M_7$ often lead to lower-than-true truncation levels, especially for larger $c_1$ and $c_2$. 
However, for $n=100$, the influence of the prior diminishes considerably, demonstrating that as the information from the likelihood grows, the prior’s effect on the posterior weakens.

\begin{table}[ht]
\centering
\scalebox{0.8}{
\begin{tabular}{|l|rrrrrr|}
\hline
  $c_2=0.5$ &   &  &   &  &   & \\
\hline
& $M_0$  & $\boldsymbol{M_1}$ & $M_2$ & $M_3$ & $M_4$ & $M_5$\\
\hline
$c_1=0.1$ & 0.01 & 0.56 & 0.32 & 0.10 & 0.01 & 0.00 \\ 
$c_1=1$ & 0.01 & 0.72 & 0.25 & 0.01 & 0.01 & 0.00 \\ 
$c_1=5 $ & 0.05 & 0.87 & 0.08 & 0.00 & 0.00 & 0.00 \\ 
\hline
& $M_0$  & ${M_1}$ & $\boldsymbol{M_2}$ & $M_3$ & $M_4$ & $M_5$\\
\hline
$c_1=0.1$ & 0.00 & 0.01 & 0.67 & 0.24 & 0.08 & 0.00 \\ 
$c_1=1$ & 0.02 & 0.01 & 0.61 & 0.30 & 0.05 & 0.01 \\ 
$c_1=5$ & 0.18 & 0.16 & 0.51 & 0.14 & 0.01 & 0.00 \\ 
  \hline
& $M_0$  & ${M_1}$ & ${M_2}$ & $\boldsymbol{M_3}$ & $M_4$ & $M_5$\\
\hline
$c_1=0.1$ & 0.01 & 0.04 & 0.00 & 0.64 & 0.28 & 0.03 \\ 
$c_1=1$ & 0.02 & 0.02 & 0.02 & 0.70 & 0.23 & 0.01 \\ 
$c_1=5$ & 0.11 & 0.16 & 0.07 & 0.56 & 0.09 & 0.01 \\ 
  \hline
& $M_0$  & ${M_1}$ & ${M_2}$ & ${M_3}$ & $\boldsymbol{M_4}$ & $M_5$\\
\hline
$c_1=0.1$  & 0.04 & 0.06 & 0.06 & 0.10 & 0.56 & 0.18 \\ 
$c_1=1$  & 0.04 & 0.05 & 0.04 & 0.14 & 0.56 & 0.17 \\ 
$c_1=5$  & 0.16 & 0.21 & 0.05 & 0.19 & 0.35 & 0.04 \\ 
\hline
& $M_0$  & ${M_1}$ & ${M_2}$ & ${M_3}$ & ${M_4}$ & $\boldsymbol{M_5}$\\
\hline
$c_1=0.1$& 0.01 & 0.05 & 0.08 & 0.31 & 0.21 & 0.34 \\ 
$c_1=1$& 0.03 & 0.07 & 0.04 & 0.22 & 0.30 & 0.34 \\ 
$c_1=5$ & 0.15 & 0.21 & 0.07 & 0.16 & 0.24 & 0.17 \\ 
   \hline
\end{tabular}
\begin{tabular}{|l|rrrrrr|}
\hline
$c_2=1$ &   &  &   &  &   & \\
\hline
& $M_0$  & $\boldsymbol{M_1}$ & $M_2$ & $M_3$ & $M_4$ & $M_5$\\
\hline
$c_1=0.1$ & 0.00 & 0.61 & 0.32 & 0.07 & 0.00 & 0.00 \\ 
$c_1=1$ & 0.00 & 0.71 & 0.26 & 0.03 & 0.00 & 0.00 \\ 
$c_1=5$ & 0.06 & 0.86 & 0.08 & 0.00 & 0.00 & 0.00 \\ 
\hline
& $M_0$  & ${M_1}$ & $\boldsymbol{M_2}$ & $M_3$ & $M_4$ & $M_5$\\
\hline
$c_1=0.1$ & 0.01 & 0.00 & 0.55 & 0.39 & 0.04 & 0.01 \\ 
$c_1=1$ & 0.00 & 0.04 & 0.73 & 0.21 & 0.02 & 0.00 \\ 
$c_1=5$ & 0.15 & 0.10 & 0.66 & 0.09 & 0.00 & 0.00 \\ 
  \hline
& $M_0$  & ${M_1}$ & $M_2$ & $\boldsymbol{M_3}$ & $M_4$ & $M_5$\\
\hline
$c_1=0.1$ & 0.02 & 0.02 & 0.00 & 0.86 & 0.09 & 0.01 \\ 
$c_1=1$ & 0.01 & 0.05 & 0.02 & 0.82 & 0.08 & 0.02 \\ 
$c_1=5$ & 0.17 & 0.13 & 0.04 & 0.58 & 0.06 & 0.02 \\ 
  \hline
& $M_0$  & ${M_1}$ & $M_2$ & $M_3$ & $\boldsymbol{M_4}$ & $M_5$\\
\hline
$c_1=0.1$ & 0.00 & 0.03 & 0.03 & 0.16 & 0.60 & 0.18 \\ 
$c_1=1$ & 0.02 & 0.06 & 0.01 & 0.12 & 0.70 & 0.09 \\ 
$c_1=5$ & 0.09 & 0.21 & 0.09 & 0.10 & 0.47 & 0.04 \\ 
  \hline
& $M_0$  & ${M_1}$ & $M_2$ & $M_3$ & $M_4$ & $\boldsymbol{M_5}$\\
\hline
$c_1=0.1$ & 0.01 & 0.05 & 0.05 & 0.25 & 0.29 & 0.35 \\ 
$c_1=1$ & 0.03 & 0.04 & 0.09 & 0.20 & 0.33 & 0.31 \\ 
$c_1=5$ & 0.18 & 0.20 & 0.08 & 0.18 & 0.29 & 0.07 \\ 
   \hline
\end{tabular}
}
\caption{Simulation study: frequencies of the selected truncation level across 100 samples with $d=6$, $n=25$, $c_1=(0.1,1,5)$ and $c_2=(0.5,1)$. Here, $M_0$ denotes the independence model, while $M_{\iota}$, $\iota=1,\ldots,5$, denotes the R-vine model truncated after tree $T_\iota$. Bold text indicates the data-generating model.}
\label{Sensitivity1}
\end{table}

\begin{table}[!ht]
\centering
\scalebox{0.8}{
\begin{tabular}{|l|rrrrrr|}
\hline
  $c_2=0.5$ &   &  &   &  &   \\
\hline
& $ M_0 $ & $\boldsymbol{M_1}$ & $M_2$ & $M_3$ & $M_4$ & $M_5$\\
\hline
$c_1=0.1$ & 0.00 & {0.81}   & 0.19  & 0.00   & 0.00  & 0.00    \\
$c_1=1$& 0.00 & {0.82}    & 0.17 &  0.01   &  0.00   & 0.00    \\
$c_1=5$ & 0.00& {0.94}   & 0.05  & 0.01    & 0.00  &  0.00     \\
\hline
& $ M_0 $ & $M_1$ & $\boldsymbol{M_2}$ & ${M_3}$ & $M_4$ & $M_5$\\
\hline
$c_1=0.1$& 0.00 & 0.00  & {0.66}  &  0.29 &  0.04 &  0.01   \\
$c_1=1$& 0.00 & 0.00  & {0.63}  &  0.33 &  0.03  & 0.01  \\
$c_1=5$ & 0.00 & 0.00  & {0.69}  &  0.25  &  0.06   & 0.00   \\
\hline
& $ M_0 $ & $M_1$ & $M_2$ & $\boldsymbol{M_3}$ & $M_4$ & $M_5$ \\
\hline
$c_1=0.1$& 0.00 & 0.00  & 0.00  & {0.79}  &  0.19  &  0.02  \\
$c_1=1$& 0.00 & 0.00  & 0.00  & {0.87}   & 0.12  &   0.01   \\
$c_1=5 $ & 0.00 & 0.00  & 0.00  & {0.87}  & 0.13   & 0.00 \\
\hline
& $ M_0 $ & $M_1$ & $M_2$ & $M_3$ & $\boldsymbol{M_4}$ & $M_5$ \\
\hline
$c_1=0.1$ & 0.00  & 0.00  & 0.00  & 0.04 &  {0.71}   & 0.25   \\
$c_1=1$ & 0.00 & 0.00  & 0.00  & 0.05 &  {0.68}   &  0.27    \\
$c_1=5 $ & 0.00 & 0.00  & 0.00  &  0.11 & {0.63}    & 0.26   \\
\hline
& $ M_0 $ & $M_1$ & $M_2$ & $M_3$ & $M_4$ & $\boldsymbol{M_5}$ \\
\hline
$c_1=0.1$ & 0.00 & 0.00  & 0.00 & 0.10   & 0.09  & {0.81}    \\
$c_1=1$ & 0.00 & 0.00  & 0.00 & 0.06   &  0.08   & {0.86}    \\
$c_1=5 $ & 0.00 & 0.00  & 0.00 & 0.15   & 0.09  &  {0.76}     \\
\hline
\end{tabular}
\begin{tabular}{|l|rrrrrr|}
\hline
 $c_2=1$ &  &   &  &   &  &   \\
\hline
& $ M_0 $ & $\boldsymbol{M_1}$ & $M_2$ & $M_3$ & $M_4$ & $M_5$\\
\hline
$c_1=0.1$ & 0.00 & {0.77}  & 0.20  &  0.02   & 0.01 & 0.00\\
$c_1=1$ & 0.00 & {0.84}  & 0.15   & 0.01   & 0.00  & 0.00   \\
$c_1=5 $ & 0.00 & {0.93}    & 0.07     & 0.00  & 0.00  & 0.00  \\
\hline
& $ M_0 $  & $M_1$ & $\boldsymbol{M_2}$ & ${M_3}$ & $M_4$ & $M_5$\\
\hline
$c_1=0.1$ & 0.00 & 0.00  & {0.78} & 0.19  & 0.03 & 0.00  \\
$c_1=1$ & 0.00 & 0.00  & {0.74} & 0.23   & 0.03  & 0.00  \\
$c_1=5 $ & 0.00 & 0.00  & {0.80}   & 0.18  &  0.02   & 0.00   \\
\hline
& $ M_0 $ &  $M_1$ & $M_2$ & $\boldsymbol{M_3}$ & $M_4$ & $M_5$ \\
\hline
$c_1=0.1$ & 0.00 & 0.00  & 0.00  & {0.84} & 0.15  & 0.01 \\
$c_1=1$ & 0.00 & 0.00  & 0.00  & {0.85} & 0.14  & 0.01    \\
$c_1=5 $ & 0.00 & 0.00  & 0.00  & {0.87}  & 0.13  & 0.00 \\
\hline
& $ M_0 $ & $M_1$ & $M_2$ & $M_3$ & $\boldsymbol{M_4}$ & $M_5$ \\
\hline
$c_1=0.1$ & 0.00 & 0.00  &  0.01   & 0.03   & {0.65}  &  0.31\\
$c_1=1$ & 0.00 & 0.00  & 0.00  & 0.03  &  {0.69}  & 0.28 \\
$c_1=5 $ & 0.00 & 0.00  & 0.00  & 0.07   & {0.70}  & 0.23  \\
\hline
& $ M_0 $ & $M_1$ & $M_2$ & $M_3$ & $M_4$ & $\boldsymbol{M_5}$ \\
\hline
$c_1=0.1$ & 0.00 & 0.00  & 0.00  & 0.18  &  0.08  &  {0.74}  \\
$c_1=1$ & 0.00 & 0.00  & 0.00  & 0.09  & 0.11   & {0.80}  \\
$c_1=5 $ & 0.00 &  0.01  & 0.01  &  0.16   & 0.07  &  {0.75}   \\
\hline
\end{tabular}
}
\caption{Simulation study: frequencies of the selected truncation level across 100 samples with $d=6$, $n=100$, $c_1=(0.1,1,5)$ and $c_2=(0.5,1)$. Here, $M_0$ denotes the independence model, while $M_\iota$, $\iota=1,\ldots,5$, denotes the R-vine model truncated after tree $T_\iota$. Bold text indicates the data-generating model.}
\label{Sensitivity2}
\end{table}

\begin{table}[ht]
\centering
\scalebox{0.7}{
\begin{tabular}{|l|rrrrrrrr|}
  \hline
$c_2=0.5$ & & & & & &  & &  \\ 
  \hline
 & $M_0$ & $\boldsymbol{M_1}$ & $M_2$ & $M_3$ & $M_4$ & $M_5$ & $M_6$ & $M_7$ \\ 
  \hline
$c_1=0.1$ & 0.00 & 0.45 & 0.29 & 0.17 & 0.09 & 0.00 & 0.00 & 0.00 \\ 
$c_1=1$ & 0.00 & 0.55 & 0.33 & 0.05 & 0.06 & 0.01 & 0.00 & 0.00 \\ 
$c_1=5$ & 0.01 & 0.88 & 0.10 & 0.01 & 0.00 & 0.00 & 0.00 & 0.00 \\ 
    \hline
 & $M_0$ & ${M_1}$ & $\boldsymbol{M_2}$ & $M_3$ & $M_4$ & $M_5$ & $M_6$ & $M_7$ \\ 
  \hline
$c_1=0.1$  & 0.15 & 0.00 & 0.41 & 0.24 & 0.13 & 0.06 & 0.01 & 0.00 \\ 
$c_1=1$ & 0.15 & 0.00 & 0.51 & 0.21 & 0.10 & 0.01 & 0.02 & 0.00 \\ 
$c_1=5$ & 0.37 & 0.00 & 0.44 & 0.17 & 0.02 & 0.00 & 0.00 & 0.00 \\ 
    \hline
 & $M_0$ & ${M_1}$ & $M_2$ & $\boldsymbol{M_3}$ & $M_4$ & $M_5$ & $M_6$ & $M_7$ \\ 
  \hline
$c_1=0.1$  & 0.07 & 0.01 & 0.00 & 0.43 & 0.36 & 0.10 & 0.03 & 0.00 \\ 
$c_1=1$& 0.09 & 0.00 & 0.03 & 0.52 & 0.23 & 0.11 & 0.02 & 0.00 \\ 
$c_1=5$ & 0.22 & 0.00 & 0.02 & 0.58 & 0.14 & 0.03 & 0.01 & 0.00 \\ 
    \hline
 & $M_0$ & ${M_1}$ & $M_2$ & $M_3$ & $\boldsymbol{M_4}$ & $M_5$ & $M_6$ & $M_7$ \\ 
  \hline
$c_1=0.1$ & 0.01 & 0.04 & 0.03 & 0.01 & 0.48 & 0.30 & 0.10 & 0.03 \\ 
$c_1=1$ & 0.03 & 0.03 & 0.09 & 0.02 & 0.55 & 0.19 & 0.09 & 0.00 \\ 
$c_1=5$ & 0.06 & 0.04 & 0.16 & 0.02 & 0.60 & 0.10 & 0.02 & 0.00 \\ 
    \hline
 & $M_0$ & ${M_1}$ & $M_2$ & $M_3$ & $M_4$ & $\boldsymbol{M_5}$ & $M_6$ & $M_7$ \\ 
  \hline
$c_1=0.1$  & 0.06 & 0.02 & 0.06 & 0.01 & 0.01 & 0.61 & 0.19 & 0.04 \\ 
$c_1=1$ & 0.13 & 0.00 & 0.07 & 0.00 & 0.00 & 0.60 & 0.18 & 0.02 \\ 
$c_1=5$ & 0.17 & 0.04 & 0.09 & 0.00 & 0.07 & 0.53 & 0.08 & 0.02 \\ 
    \hline
 & $M_0$ & ${M_1}$ & $M_2$ & $M_3$ & $M_4$ & $M_5$ & $\boldsymbol{M_6}$ & $M_7$ \\ 
  \hline
$c_1=0.1$  & 0.08 & 0.03 & 0.05 & 0.00 & 0.02 & 0.31 & 0.34 & 0.17 \\ 
$c_1=1$ & 0.09 & 0.05 & 0.09 & 0.00 & 0.05 & 0.23 & 0.29 & 0.20 \\ 
$c_1=5$ & 0.20 & 0.06 & 0.15 & 0.01 & 0.06 & 0.28 & 0.19 & 0.05 \\ 
    \hline
 & $M_0$ & ${M_1}$ & $M_2$ & $M_3$ & $M_4$ & $M_5$ & $M_6$ & $\boldsymbol{M_7}$ \\ 
  \hline
$c_1=0.1$  & 0.04 & 0.02 & 0.08 & 0.01 & 0.02 & 0.44 & 0.13 & 0.26 \\ 
$c_1=1$ & 0.05 & 0.05 & 0.08 & 0.00 & 0.06 & 0.44 & 0.18 & 0.14 \\ 
$c_1=5$ & 0.22 & 0.08 & 0.11 & 0.01 & 0.07 & 0.30 & 0.10 & 0.11 \\ 
   \hline
   \end{tabular}
   \begin{tabular}{|l|rrrrrrrr|}
  \hline
$c_2=1$ & & & & & &  & &  \\ 
  \hline
 & $M_0$ & $\boldsymbol{M_1}$ & $M_2$ & $M_3$ & $M_4$ & $M_5$ & $M_6$ & $M_7$ \\ 
  \hline
$c_1=0.1$ & 0.00 & 0.44 & 0.29 & 0.17 & 0.09 & 0.01 & 0.00 & 0.00 \\ 
$c_1=1$ & 0.00 & 0.60 & 0.26 & 0.08 & 0.05 & 0.01 & 0.00 & 0.00 \\ 
$c_1=5$ & 0.05 & 0.81 & 0.12 & 0.02 & 0.00 & 0.00 & 0.00 & 0.00 \\ 
    \hline
 & $M_0$ & $M_1$ & $\boldsymbol{M_2}$ & $M_3$ & $M_4$ & $M_5$ & $M_6$ & $M_7$ \\ 
  \hline
$c_1=0.1$ & 0.07 & 0.00 & 0.46 & 0.27 & 0.16 & 0.04 & 0.00 & 0.00 \\ 
$c_1=1$  & 0.10 & 0.00 & 0.54 & 0.26 & 0.10 & 0.00 & 0.00 & 0.00 \\ 
$c_1=5$ & 0.31 & 0.00 & 0.45 & 0.21 & 0.03 & 0.00 & 0.00 & 0.00 \\ 
    \hline
 & $M_0$ & $M_1$ & $M_2$ & $\boldsymbol{M_3}$ & $M_4$ & $M_5$ & $M_6$ & $M_7$ \\ 
  \hline
$c_1=0.1$& 0.07 & 0.01 & 0.00 & 0.43 & 0.36 & 0.10 & 0.03 & 0.00 \\ 
$c_1=1$  & 0.09 & 0.00 & 0.03 & 0.52 & 0.23 & 0.11 & 0.02 & 0.00 \\ 
$c_1=5$ & 0.22 & 0.00 & 0.02 & 0.58 & 0.14 & 0.03 & 0.01 & 0.00 \\ 
    \hline
 & $M_0$ & $M_1$ & $M_2$ & $M_3$ & $\boldsymbol{M_4}$ & $M_5$ & $M_6$ & $M_7$ \\ 
  \hline
$c_1=0.1$& 0.03 & 0.01 & 0.06 & 0.00 & 0.61 & 0.22 & 0.07 & 0.00 \\ 
$c_1=1$   & 0.03 & 0.08 & 0.05 & 0.00 & 0.66 & 0.08 & 0.09 & 0.01 \\ 
$c_1=5$ & 0.13 & 0.03 & 0.17 & 0.00 & 0.60 & 0.06 & 0.01 & 0.00 \\ 
    \hline
 & $M_0$ & $M_1$ & $M_2$ & $M_3$ & $M_4$ & $\boldsymbol{M_5}$ & $M_6$ & $M_7$ \\ 
  \hline
$c_1=0.1$ & 0.05 & 0.05 & 0.04 & 0.00 & 0.06 & 0.63 & 0.14 & 0.03 \\ 
$c_1=1$  & 0.10 & 0.05 & 0.06 & 0.00 & 0.05 & 0.66 & 0.07 & 0.01 \\ 
$c_1=5$ & 0.15 & 0.05 & 0.12 & 0.00 & 0.01 & 0.62 & 0.04 & 0.01 \\ 
    \hline
 & $M_0$ & $M_1$ & $M_2$ & $M_3$ & $M_4$ & $M_5$ & $\boldsymbol{M_6}$ & $M_7$ \\ 
  \hline
$c_1=0.1$& 0.05 & 0.05 & 0.08 & 0.00 & 0.02 & 0.35 & 0.37 & 0.08 \\ 
$c_1=1$   & 0.06 & 0.01 & 0.09 & 0.01 & 0.11 & 0.27 & 0.35 & 0.10 \\ 
$c_1=5$ & 0.17 & 0.09 & 0.15 & 0.01 & 0.05 & 0.30 & 0.17 & 0.06 \\ 
    \hline
 & $M_0$ & $M_1$ & $M_2$ & $M_3$ & $M_4$ & $M_5$ & $M_6$ & $\boldsymbol{M_7}$ \\ 
  \hline
$c_1=0.1$ & 0.04 & 0.05 & 0.05 & 0.00 & 0.02 & 0.39 & 0.28 & 0.17 \\ 
$c_1=1$   & 0.11 & 0.06 & 0.06 & 0.01 & 0.01 & 0.41 & 0.21 & 0.13 \\ 
$c_1=5$   & 0.12 & 0.05 & 0.17 & 0.00 & 0.14 & 0.37 & 0.10 & 0.05 \\ 
   \hline
\end{tabular}}
\caption{Simulation study: frequencies of the selected truncation level across 100 samples with $d=8$, $n=25$, $c_1=(0.1,1,5)$ and $c_2=(0.5,1)$. Here, $M_0$ denotes the independence model, while $M_\iota$, $\iota=1,\ldots,7$, denotes the R-vine model truncated after tree $T_\iota$. Bold text indicates the data-generating model.}
\label{Sensitivity3}
\end{table}

\begin{table}[ht]
\centering
\scalebox{0.7}{
\begin{tabular}{|r|rrrrrrrr|}
  \hline
 $c_2=0.5$& $M_0$ & $\boldsymbol{M_1}$ & $M_2$ & $M_3$ & $M_4$ & $M_5$ & $M_6$ & $M_7$ \\ 
  \hline
$c_1=0.1$  & 0.00 & 0.76 & 0.15 & 0.07 & 0.02 & 0.00 & 0.00 & 0.00 \\ 
$c_1=1$ & 0.00 & 0.78 & 0.18 & 0.04 & 0.00 & 0.00 & 0.00 & 0.00 \\ 
$c_1=5$ & 0.00 & 0.91 & 0.08 & 0.01 & 0.00 & 0.00 & 0.00 & 0.00 \\ 
    \hline
 & $M_0$ & ${M_1}$ & $\boldsymbol{M_2}$ & $M_3$ & $M_4$ & $M_5$ & $M_6$ & $M_7$ \\ 
  \hline
$c_1=0.1$  & 0.05 & 0.00 & 0.59 & 0.22 & 0.13 & 0.01 & 0.00 & 0.00 \\ 
$c_1=1$ & 0.14 & 0.00 & 0.72 & 0.10 & 0.02 & 0.02 & 0.00 & 0.00 \\ 
$c_1=5$ & 0.24 & 0.00 & 0.64 & 0.10 & 0.02 & 0.00 & 0.00 & 0.00 \\ 
    \hline
 & $M_0$ & ${M_1}$ & $M_2$ & $\boldsymbol{M_3}$ & $M_4$ & $M_5$ & $M_6$ & $M_7$ \\ 
  \hline
$c_1=0.1$ & 0.01 & 0.00 & 0.00 & 0.65 & 0.30 & 0.03 & 0.00 & 0.01 \\ 
$c_1=1$  & 0.02 & 0.00 & 0.00 & 0.65 & 0.27 & 0.06 & 0.00 & 0.00 \\ 
$c_1=5$ & 0.01 & 0.00 & 0.00 & 0.78 & 0.19 & 0.01 & 0.01 & 0.00 \\ 
    \hline
 & $M_0$ & ${M_1}$ & $M_2$ & $M_3$ & $\boldsymbol{M_4}$ & $M_5$ & $M_6$ & $M_7$ \\ 
  \hline
$c_1=0.1$  & 0.00 & 0.00 & 0.00 & 0.00 & 0.64 & 0.27 & 0.07 & 0.02 \\ 
$c_1=1$ & 0.00 & 0.00 & 0.00 & 0.00 & 0.70 & 0.22 & 0.05 & 0.03 \\ 
$c_1=5$  & 0.00 & 0.00 & 0.05 & 0.00 & 0.64 & 0.20 & 0.09 & 0.02 \\ 
    \hline
 & $M_0$ & ${M_1}$ & $M_2$ & $M_3$ & $M_4$ & $\boldsymbol{M_5}$ & $M_6$ & $M_7$ \\ 
  \hline
$c_1=0.1$  & 0.00 & 0.00 & 0.00 & 0.00 & 0.00 & 0.61 & 0.31 & 0.08 \\ 
$c_1=1$ & 0.02 & 0.00 & 0.00 & 0.00 & 0.00 & 0.66 & 0.26 & 0.06 \\ 
$c_1=5$  & 0.01 & 0.00 & 0.01 & 0.00 & 0.00 & 0.68 & 0.27 & 0.03 \\ 
    \hline
 & $M_0$ & ${M_1}$ & $M_2$ & $M_3$ & $M_4$ & $M_5$ & $\boldsymbol{M_6}$ & $M_7$ \\ 
  \hline
$c_1=0.1$  & 0.01 & 0.00 & 0.00 & 0.00 & 0.00 & 0.14 & 0.57 & 0.28 \\ 
$c_1=1$ & 0.01 & 0.00 & 0.01 & 0.00 & 0.00 & 0.08 & 0.66 & 0.24 \\ 
$c_1=5$ & 0.03 & 0.00 & 0.01 & 0.00 & 0.00 & 0.23 & 0.44 & 0.29 \\ 
    \hline
 & $M_0$ & ${M_1}$ & $M_2$ & $M_3$ & $M_4$ & $M_5$ & $M_6$ & $\boldsymbol{M_7}$ \\ 
  \hline
$c_1=0.1$ & 0.01 & 0.01 & 0.00 & 0.00 & 0.00 & 0.02 & 0.23 & 0.73 \\ 
$c_1=1$ & 0.00 & 0.00 & 0.00 & 0.00 & 0.00 & 0.01 & 0.26 & 0.73 \\ 
$c_1=5$ & 0.01 & 0.00 & 0.01 & 0.00 & 0.00 & 0.02 & 0.33 & 0.63 \\ 
   \hline
\end{tabular}
\begin{tabular}{|r|rrrrrrrr|}
  \hline
 $c_2=1$ & $M_0$ & $\boldsymbol{M_1}$ & $M_2$ & $M_3$ & $M_4$ & $M_5$ & $M_6$ & $M_7$ \\ 
  \hline
$c_1=0.1$ & 0.00 & 0.74 & 0.17 & 0.06 & 0.02 & 0.01 & 0.00 & 0.00 \\ 
$c_1=1$  & 0.00 & 0.77 & 0.18 & 0.04 & 0.01 & 0.00 & 0.00 & 0.00 \\ 
$c_1=5$  & 0.00 & 0.92 & 0.06 & 0.02 & 0.00 & 0.00 & 0.00 & 0.00 \\ 
  \hline
 & $M_0$ & $M_1$ & $\boldsymbol{M_2}$ & $M_3$ & $M_4$ & $M_5$ & $M_6$ & $M_7$ \\ 
  \hline
$c_1=0.1$  & 0.06 & 0.00 & 0.72 & 0.14 & 0.08 & 0.00 & 0.00 & 0.00 \\ 
$c_1=1$& 0.10 & 0.00 & 0.74 & 0.12 & 0.04 & 0.00 & 0.00 & 0.00 \\ 
$c_1=5$  & 0.24 & 0.00 & 0.69 & 0.05 & 0.02 & 0.00 & 0.00 & 0.00 \\ 
  \hline
 & $M_0$ & $M_1$ & $M_2$ & $\boldsymbol{M_3}$ & $M_4$ & $M_5$ & $M_6$ & $M_7$ \\ 
  \hline
$c_1=0.1$ & 0.00 & 0.00 & 0.00 & 0.68 & 0.28 & 0.04 & 0.00 & 0.00 \\ 
$c_1=1$ & 0.04 & 0.00 & 0.00 & 0.69 & 0.25 & 0.02 & 0.00 & 0.00 \\ 
$c_1=5$ & 0.06 & 0.00 & 0.00 & 0.74 & 0.19 & 0.01 & 0.00 & 0.00 \\ 
  \hline
 & $M_0$ & $M_1$ & $M_2$ & $M_3$ & $\boldsymbol{M_4}$ & $M_5$ & $M_6$ & $M_7$ \\ 
  \hline
$c_1=0.1$  & 0.01 & 0.00 & 0.00 & 0.00 & 0.72 & 0.18 & 0.09 & 0.00 \\ 
$c_1=1$ & 0.00 & 0.00 & 0.02 & 0.00 & 0.75 & 0.18 & 0.04 & 0.01 \\ 
$c_1=5$ & 0.00 & 0.00 & 0.01 & 0.00 & 0.73 & 0.21 & 0.04 & 0.01 \\
  \hline
 & $M_0$ & $M_1$ & $M_2$ & $M_3$ & $M_4$ & $\boldsymbol{M_5}$ & $M_6$ & $M_7$ \\ 
  \hline
$c_1=0.1$ & 0.00 & 0.00 & 0.00 & 0.00 & 0.00 & 0.69 & 0.27 & 0.04 \\ 
$c_1=1$ & 0.00 & 0.00 & 0.00 & 0.00 & 0.00 & 0.67 & 0.31 & 0.02 \\ 
$c_1=5$  & 0.01 & 0.00 & 0.00 & 0.00 & 0.00 & 0.72 & 0.25 & 0.02 \\ 
  \hline
 & $M_0$ & $M_1$ & $M_2$ & $M_3$ & $M_4$ & $M_5$ & $\boldsymbol{M_6}$ & $M_7$ \\ 
  \hline
$c_1=0.1$ & 0.01 & 0.00 & 0.00 & 0.00 & 0.00 & 0.18 & 0.65 & 0.16 \\ 
$c_1=1$ & 0.01 & 0.00 & 0.00 & 0.00 & 0.00 & 0.14 & 0.53 & 0.32 \\ 
$c_1=5$  & 0.04 & 0.01 & 0.03 & 0.00 & 0.00 & 0.28 & 0.47 & 0.17 \\ 
  \hline
 & $M_0$ & $M_1$ & $M_2$ & $M_3$ & $M_4$ & $M_5$ & $M_6$ & $\boldsymbol{M_7}$ \\ 
  \hline
$c_1=0.1$  & 0.00 & 0.00 & 0.01 & 0.00 & 0.00 & 0.06 & 0.25 & 0.68 \\ 
$c_1=1$& 0.02 & 0.01 & 0.00 & 0.00 & 0.00 & 0.06 & 0.24 & 0.67 \\ 
$c_1=5$ & 0.01 & 0.01 & 0.01 & 0.00 & 0.01 & 0.08 & 0.38 & 0.50 \\ 
   \hline
\end{tabular}
}
\caption{Simulation study: frequencies of the selected truncation level across 100 samples with $d=8$, $n=100$, $c_1=(0.1,1,5)$ and $c_2=(0.5,1)$. Here, $M_0$ denotes the independence model, while $M_\iota$, $\iota=1,\ldots,7$, denotes the R-vine model truncated after tree $T_\iota$. Bold text indicates the data-generating model.}
\label{Sensitivity4}
\end{table}

\section{Application to Portfolio Asset Returns}

In order to test our approach on real data, we employed a dataset previously analysed by \cite{gruber2015sequential,gruber2018bayesian}. The dataset comprises the adjusted daily closing prices of nine iShares ETFs, where $\delta = 1:9$,
spanning the period from January 2013 to June 2014. The training set includes 252 data points from
January to December 2013 ($t = 1:252$), while the test set contains 124 data points from January
to June 2014 ($t = 253:376$). The nine ETFs represent a diversified portfolio across several asset
classes that can be easily replicated by individual investors. Three of the ETFs focus on U.S. equities ($\delta = 1, 2, 3$), two invest in U.S. treasuries ($\delta = 4, 5$), two are based on U.S. real estate through real estate investment trusts (REITs, $\delta = 6, 7$), and two are commodity trusts that focus on gold and silver ($\delta = 8, 9$). For more details about the assets included in the dataset, please see \cite{gruber2015sequential,gruber2018bayesian}.

In order to model this dataset and analyse the dependence structure amongst the variables, we adopt the two-step approach of \cite{JoeXu1996}, estimating the marginals first, and, successively, estimating an R-vine copula.
Focusing on the margins first, the daily log-returns $y_{\delta,t}$, for $t = 1, 2, \ldots, n$, of each series ($\delta = 1:9$) are modeled using a dynamic linear model (DLM) with variance discounting (see \cite{west2006bayesian}). The DLM is a fully Bayesian time series approach, featuring closed-form posterior and forecast distributions, with parameters updated in real-time. The updating formulas used in the present application are adapted from Table~10.4 in \cite{west2006bayesian}. 
For further details on the calibration and model selection used to model the marginal distributions, see \cite{gruber2015sequential, gruber2018bayesian}. 

In order to model the dependence amongst the financial assets, we then estimated an R-vine copula, comparing our proposed approach with the gold standard, which is the sequential pair copula selection and estimation approach by \cite{dissmann2013selecting}.
In Tables~\ref{freqRes} and \ref{ourModel}, we present the results in terms of model selection. From the results, it is clear that the optimal model selected by our proposed method is significantly more parsimonious than the one selected by the traditional method by \cite{dissmann2013selecting}.
For the $9$-dimensional R-vine, the selected truncation level is the fifth, with many copula families being the independence copula. 

\begin{figure}[!ht]
    \centering
    \includegraphics[width=1\linewidth]{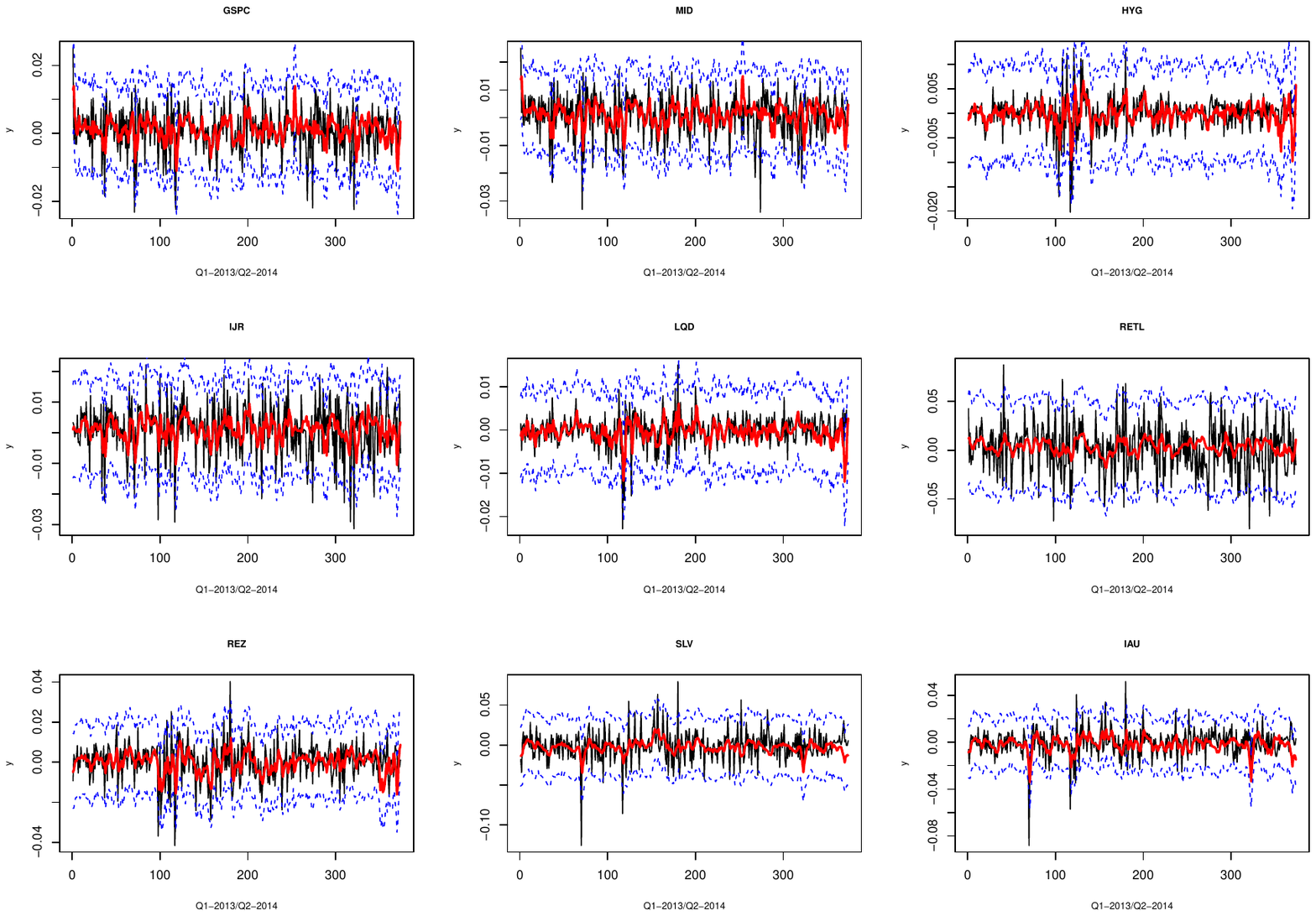}
    \caption{Forecasts of the adjusted daily closing prices for the $9$-dimensional iShares ETFs dataset. In each panel, the time horizon is on the $x$-axis, while the asset prices are on the $y$-axis. The panel indicates each financial asset.
    The red solid lines show the expected values of the predictive distribution, the dashed blue lines show the credible intervals, calculated using $q_{0.025}$ and $q_{0.975}$, while the black solid lines show the observed values.}
    \label{predictions}
\end{figure}

Figure~\ref{predictions} shows the predictive performance of the R-vine model we selected with our proposed approach.
More precisely, Figure~\ref{predictions} depicts the forecasted values of the adjusted daily closing prices for the $9$-dimensional iShares ETFs dataset. In each panel, the time horizon is on the horizontal axis, while the asset prices are on the vertical axis. The panel indicates each financial asset.
The red solid lines show the expected values of the predictive distribution, the dashed blue lines show the credible intervals, calculated using $q_{0.025}$ and $q_{0.975}$, while the black solid lines show the observed values.
The results demonstrate that the R-vine selected by the proposed approach provides a good performance in terms of in-sample predictions, especially for assets $\delta = 1,2,3,4,5,7$.

Finally, we carried out a comparison in terms of out-of-sample forecasts between our proposed approach and the one by \cite{dissmann2013selecting}.
Table \ref{RMSE} reports the root mean squared error (RMSE) calculated on the out-of-sample predictions for both models \citep{merhav1998universal}. 
The financial assets are listed in the columns, while the methods are listed in the rows.
Comparing the results we note that the RMSEs or each asset are virtually identical for both approaches,
showing that the out-of-sample predictive performance of our method is comparable to that of the traditional state-of-the-art approach. 
Therefore, although our approach selects a more parsimonious, and thus less flexible, R-vine model, it achieves predictive performance that is essentially equivalent to that of the model proposed by \cite{dissmann2013selecting}.

\begin{table}[!ht]
\centering
\scalebox{0.8}{
\begin{tabular}{|l|l|l|l|l|}
\hline
Tree $T_1$ & Tree $T_2$ & Tree $T_3$ & Tree $T_4$ & Tree $T_5$ \\ 
\hline
$c_{1,6}$ G(0.56) & $c_{2,6;1}$ G(0.26) & $c_{4,6;2,1}$ RG90(-0.15) & $c_{7,6;4,2,1}$ SC(0.01) & $c_{3,6;7,4,2,1}$ G(-0.04) \\ 
\hline
$c_{2,4}$ SG(0.79) & $c_{1,4;2}$ G(0.09) & $c_{7,4;1,2}$ G(-0.07) & $c_{3,4;7,1,2}$ C(0.03) & $c_{5,4;3,7,1,2}$ G(-0.08) \\ 
\hline
$c_{1,7}$ G(0.43) & $c_{2,7;1}$ G(0.11) & $c_{3,7;2,1}$ G(0.17) & $c_{5,7;3,2,1}$ G(0.24) & $c_{9,7;5,3,2,1}$ SC(0.06) \\ 
\hline
$c_{2,1}$ G(0.77) & $c_{3,1;2}$ SG(0.09) & $c_{5,1;3,2}$ RG270(-0.10) & $c_{9,1;5,3,2}$ RG270(-0.05) & $c_{8,1;9,5,3,2}$ SG(0.02) \\ 
\hline
$c_{3,2}$ G(0.34) & $c_{5,2;3}$ G(-0.22) & $c_{9,2;5,3}$ G(0.14) & $c_{8,2;9,5,3}$ G(0.10) & \\ 
\hline
$c_{5,3}$ G(0.38) & $c_{9,3;5}$ C(0.02) & $c_{8,3;9,5}$ RC90(-0.02) & & \\ 
\hline
$c_{9,5}$ G(0.22) & $c_{8,5;9}$ RG90(-0.06) & & & \\ 
\hline
$c_{9,8}$ G(0.77) & & & & \\ 
\hline
Tree $T_6$ & Tree $T_7$ & Tree $T_8$ & & \\ 
\hline
$c_{5,6;3,7,4,2,1}$ RG90(-0.02) & $c_{9,6;5,3,7,4,2,1}$ SG(0.15) & $c_{8,6;9,5,3,7,4,2,1}$ SC(0.15) & & \\ 
\hline
$c_{9,4;5,3,7,1,2}$ RG90(-0.04) & $c_{8,4;9,5,3,7,1,2}$ G(-0.06) & & & \\ 
\hline
$c_{8,7;9,5,3,2,1}$ SG(0.07) & & & & \\ 
\hline
\end{tabular}
}
\caption{R-vine copula model selected with the sequential method by \cite{dissmann2013selecting} for the $9$-dimensional iShares
ETFs dataset. Kendall's $\tau$ parameters for each pair copula are in brackets.}
\label{freqRes}
\end{table}

\begin{table}[!ht]
\centering
\begin{tabular}{|l|l|l|l|l|}
\hline
Tree $T_1$            & Tree $T_2$            & Tree $T_3$            & Tree $T_4$            & Tree $T_5$ \\ \hline
$c_{1,6}$ G(0.26)     & $c_{2,6;1}$ G(0.60)   & $c_{4,6;2,1}$ I       & $c_{7,6;4,2,1}$ I     & $c_{3,6;7,4,2,1}$ I   \\ \hline
$c_{2,4}$ N(0.79)     & $c_{1,4;2}$ N(0.21)   & $c_{7,4;1,2}$ I       & $c_{3,4;7,1,2}$ I     & $c_{5,4;3,7,1,2}$ I   \\ \hline
$c_{1,7}$ N(0.43)     & $c_{2,7;1}$ I         & $c_{3,7;2,1}$ N(0.17) & $c_{5,7;3,2,1}$ G(0.22) & $c_{9,7;5,3,2,1}$ I   \\ \hline
$c_{2,1}$ N(0.56)     & $c_{3,1;2}$ I         & $c_{5,1;3,2}$ I       & $c_{9,1;5,3,2}$ G(0.30) & $c_{8,1;9,5,3,2}$ N(0.30) \\ \hline
$c_{3,2}$ G(0.34)     & $c_{5,2;3}$ N(-0.22)  & $c_{9,2;5,3}$ I       & $c_{8,2;9,5,3}$ I     & \\ \hline
$c_{5,3}$ N(0.38)     & $c_{9,3;5}$ I         & $c_{8,3;9,5}$ I       & & \\ \hline
$c_{9,5}$ I           & $c_{8,5;9}$ N(-0.20)  & & & \\ \hline
$c_{9,8}$ G(0.79)     & & & & \\ \hline
\end{tabular}
\caption{R-vine copula model selected with our proposed approach for the $9$-dimensional iShares
ETFs dataset. Kendall's $\tau$ parameters for each pair copula are in brackets.}
\label{ourModel}
\end{table}

\begin{table}[!ht]
\centering
\begin{tabular}{|l|rrrrrrrrr|}
  \hline
 & GSPC & MID & HYG & IJR & LQD & RETL & REZ & SLV & IAU \\ 
  \hline
\cite{dissmann2013selecting} & 0.01 & 0.01 & 0.00 & 0.01 & 0.00 & 0.00 & 0.00 & 0.03 & 0.02 \\ 
  \textbf{Proposed Method} & 0.01 & 0.01 & 0.00 & 0.01 & 0.00 & 0.00 & 0.00 & 0.03 & 0.02 \\ 
   \hline
\end{tabular}
\caption{Root mean squared error (RMSE) of the out-of-sample predictions for each asset (columns) for Dissman's method (first row) and the proposed method (second row).}
\label{RMSE}
\end{table}


\section{Concluding Remarks}

In this paper, we propose a Bayesian framework for model selection and estimation of R-vine copula models by using a loss-based perspective. The approach jointly addresses the selection of pair-copula families and the structure while accounting for model uncertainty through sparsity-inducing loss-based priors. 

The performance of the proposed methodology was investigated through an extensive simulation study. The results showed that the procedure is able to accurately recover both the underlying copula families and the true vine structure. 

The analysis of the financial data example further demonstrated the power of the proposed approach. The selected model achieved predictive performance comparable to that obtained using the existing state-of-the-art method, while relying on a more parsimonious R-vine specification. This result highlights the ability of the proposed framework to balance model complexity and predictive accuracy.

Further work include extensions to dynamic and time-varying vine structures, as well as conditional copula models, where copula parameters vary according to the values of covariates.


\newpage

\appendix
\section{Appendix: Loss-Based Priors for Gaussian Vine Copulas}\label{se:KL}

In this Section, we investigate the behaviour of the Kullback--Leibler (KL) divergence, and consequently the associated loss of information, for vine-based models in which all pair-copulas belong to the Gaussian family and all marginal distributions are Gaussian with zero mean. This assumption is made without loss of generality. 
Since we assume that $\varphi$ is fixed and corresponds to the Gaussian family, according to the notation introduced in Section~\ref{sec:methodology}, our model notation simplifies to $M_{j} = M_{j|\varphi}$.

\subsection{Loss in Information in the Bivariate Gaussian Case}\label{sc:KLnormalsimple}

Let us consider the random variables ($X_1, X_2$) in the bivariate case ($d=2$), where only two possible models are available. 
Model $M_0$ assumes independence between the two random variables; hence, from Eq.~\eqref{eq:joint} the model can be expressed as:
$$M_0: \{ f_1 f_2 \}$$
where, to simplify the notation, we set $f_1(x_1)=f_1$ and $f_2(x_2)=f_2$. In addition, we assume that $f_{\iota}\sim N(0,\sigma_{\iota}^2)$, for ${\iota}=1,2$. 
In contrast, model $M_1$ assumes dependence, described by the Gaussian copula density with dependence parameter $\rho$:
$$c(u_1,u_2) = \frac{1}{\sqrt{1-\rho^2}}\exp\left\{\frac{2\rho\Phi^{-1}(u_1)\Phi^{-1}(u_2)-\rho^2\left(\Phi^{-1}(u_1)^2+\Phi^{-1}(u_2)^2\right)}{2(1-\rho^2)}\right\}.$$
\noindent
Therefore, we set
$$M_1: \{ c(u_1,u_2) \, f_1f_2 \}.$$
Recall that, in general, if the random variable $W$ has distribution function $F(w)$, then for $\omega\in(0,1)$,
$$\int F^{-1}(\omega)\,d\omega = \mathbb{E}(W) \qquad \mbox{and} \qquad \int\Phi^{-1}(\omega)^2\,d\omega = \mathbb{E}(W^2).$$
Thus, considering that the copula density for model $M_0$ is the independence copula, the KL divergence between the two models under consideration is
$$D_{KL}(M_0\|M_1) = \int f_1f_2\log\left(\frac{f_1f_2}{f_1f_2\,c(u_1,u_2)}\right)\,du_1\,du_2 = \int\log\left(\frac{1}{c(u_1,u_2)}\right)\,du_1\,du_2,$$
since the KL divergence between two distributions is equal to the KL between the copulas, provided the marginals are the same. Thus,
\begin{eqnarray}\label{eq:KLm0m1}
D_{KL}(M_0\|M_1) &=& \mathbb{E}_{M_0}\left[\frac{1}{c(u_1,u_2)}\right] \nonumber \\
&=& -\mathbb{E}_{M_0}\left[c(u_1,u_2)\right] \nonumber \\
&=& -\int\left[\frac{1}{\sqrt{1-\rho^2}}\exp\left\{\frac{2\rho\Phi^{-1}(u_1)\Phi^{-1}(u_2)-\rho^2\left(\Phi^{-1}(u_1)^2+\Phi^{-1}(u_2)^2\right)}{2(1-\rho^2)}\right\}\right]\,du_1\,du_2 \nonumber \\
&=& \log\sqrt{1-\rho^2}\int\,du_1\,du_2 - \frac{1}{2(1-\rho^2)}\left[2\rho \mathbb{E}(X_1) \mathbb{E}(X_2) -\rho^2\left(\sigma_1^2+\mathbb{E}(X_1)+\sigma_2^2+\mathbb{E}(X_2)\right)\right] \nonumber \\
&=& \log\sqrt{1-\rho^2} + \frac{\rho^2(\sigma_1^2+\sigma_2^2)}{2(1-\rho^2)}.
\end{eqnarray}

\noindent
We note that
$$
\lim_{\rho\rightarrow 0} D_{KL}(M_0\|M_1)= 0 \hspace{1cm} \mbox{and} \hspace{1cm} \lim_{\rho\rightarrow\pm 1} D_{KL}(M_0\|M_1)= +\infty.
$$
From these results, we conclude that the KL divergence depends on the strength of dependence between the two random variables $(X_1, X_2)$ and that its value increases as the dependence becomes stronger. 

Finally, the prior mass for model $M_0$, based on the loss in information only, can be obtained by minimizing $\eqref{eq:KLm0m1}$ with respect to $\rho$, which leads to $P(M_0)\propto\exp(0)$.

\subsection{Loss in Information in the Multivariate Gaussian Case} \label{sc:KLnormalgeneral}

In this Section, we extend the result obtained in Section~\ref{sc:KLnormalsimple} to a $d$-dimensional D-vine structure with all Gaussian pair-copulas, and we calculate the corresponding loss in information.

Let us consider the random variables ($X_1, \ldots, X_d$) and the corresponding $d$-dimensional D-vine with ordering $X_1 - X_2 - \ldots - X_d$.
We assume that the models available in this case are given by all possible truncations, resulting in models $M_0$ to $M_{d-1}$, as listed below
\begin{eqnarray*}
M_0: && \left\{f_1\cdots f_d\right\} \\
M_1: && \left\{c_{1,2}c_{2,3}\cdots c_{d-1,d} \, \times f_1 \cdots f_d  \right\} \\
\cdots && \cdots \\
M_{d-1}: && \left\{ c_{1,2}c_{2,3}\cdots c_{d-1,d} \cdots \,c_{1,d|2,\ldots,d-1}  \, f_1\cdots f_d\right\}.
\end{eqnarray*}

The proposed approach assigns prior probability mass to each model $M_j$, for $j=0,1,\ldots,d-1$, based on the minimum KL divergence between $M_j$ and its closest competing model, denoted by $M_{j'}$. For models $M_j$, with $j=0,\ldots,d-2$, the closest model is always $M_{j+1}$, since the two models differ only through the additional copula component included in $M_{j+1}$. In this case, the minimum KL divergence between $M_j$ and $M_{j+1}$, with respect to the parameter present only in $M_{j+1}$, is equal to zero. Indeed, this is achieved by setting the corresponding correlation parameter to zero, thereby recovering model $M_j$.
The situation is different for the ``saturated'' model, $M_{d-1}$. Since there is no more complex model to which it can be compared, the prior mass assigned to $M_{d-1}$ is based on a strictly positive expected minimum KL divergence. Consequently, this prior mass depends on the uncertainty associated with the parameter(s) that are present in $M_{d-1}$ but absent from $M_{d-2}$.
Therefore, the prior mass based exclusively on the loss in information is
\begin{equation}\label{eq:prior_gauss_dvine}
    P(M_j)\propto 
\begin{cases}
    1,& \text{if } j=1,\ldots,d-2\\
    \exp\left[\int_{[-1,1]^d}\left\{\int_{[0,1]^d}M_{d-1}\log\left(c_{1,d|2,\ldots,d-1}\right)\,d\boldsymbol{u}\right\}\pi(\pmb\rho)\,d\pmb\rho\right],              & \text{if } j=d-1,
\end{cases}
\end{equation}
where $\pi(\pmb\rho)$ is a suitable prior distribution for the correlation coefficients in the deepest model. The prior mass on model $M_{d-1}$ will depend on the aforementioned prior for $\pmb\rho$.

In order to understand the behaviour of the prior mass in Eq.~\eqref{eq:prior_gauss_dvine}, particularly for $j=d-1$, we calculated the expected KL divergence for various dimensions, such as $d=3,\ldots,10,20$, where $\pi(\pmb\rho)$ is a $d$-dimensional uniform on the space $(-1,1)$. 
Table~\ref{tb:minimKLnormal} shows that the expected minimum KL divergence between the most complex model and the nearest one appears to be unaffected by the dimensionality of the problem.

\begin{table}[!ht]
\centering
\begin{tabular}{cc}
\hline 
$d$ & $\mathbb{E}\left\{D_{KL}(M_{d-1}\|M_{d-2})\right\}$ \\ 
\hline 
3 & 0.3082280 \\ 
4 & 0.3044187 \\ 
5 & 0.3051971 \\ 
6 & 0.3112703 \\ 
7 & 0.3079909 \\ 
8 & 0.3075988 \\ 
9 & 0.3056916 \\ 
10 & 0.3103587 \\
20 & 0.3032913 \\
\hline 
\end{tabular}
\caption{Expected minimum Kullback--Leibler divergence between the deepest model $(M_{d-1})$ and the nearest one $(M_{d-2})$.}
\label{tb:minimKLnormal}
\end{table}

\newpage

\bibliographystyle{apalike}
\bibliography{priors_vines}

\end{document}